\documentclass{ieeeaccess}
\usepackage{cite}
\usepackage{amsmath,amssymb,amsfonts}
\usepackage{algorithmic}
\usepackage{graphicx}
\usepackage{textcomp}
\usepackage{tabu}
\usepackage{multicol}
\usepackage{color,soul}
\usepackage{lineno,hyperref}
\usepackage{algorithm}
\usepackage{booktabs}
\usepackage{optidef}

\def\BibTeX{{\rm B\kern-.05em{\sc i\kern-.025em b}\kern-.08em
    T\kern-.1667em\lower.7ex\hbox{E}\kern-.125emX}}

\begin{document}
\history{Date of publication xxxx 00, 0000, date of current version xxxx 00, 0000.}
\doi{10.1109/ACCESS.2017.DOI}

\title{Energy-cost aware off-grid base stations with IoT devices for developing a green heterogeneous network}
\author{\uppercase{Khondoker Ziaul Islam}\authorrefmark{1,2,3}, \IEEEmembership{Graduate Student Member, IEEE},
\uppercase{Md. Sanwar Hossain}\authorrefmark{3},\IEEEmembership{Graduate Student Member, IEEE}, \uppercase{B.M. Ruhul Amin}\authorrefmark{4}, \IEEEmembership{Graduate Student Member, IEEE}, and \uppercase{Ferdous Sohel}\authorrefmark{1,2}, \IEEEmembership{Senior Member, IEEE}}
\address[1]{Discipline of Information Technology, Murdoch University, 90 South Street, Murdoch, WA 6150, Australia}
\address[2]{Centre for Crop and Food Innovation, Food Futures Institute, Murdoch University, Murdoch, WA 6150, Australia}
\address[3]{Department of Electrical and Electronic Engineering, Bangladesh University of Business and Technology, Mirpur, Dhaka-1216, Bangladesh}
\address[4]{School of Engineering, Macquarie University, Balaclava Road, Macquarie Park, NSW 2109, Australia}

\markboth
{Islam \headeretal: Energy-cost aware off-grid base stations with IoT devices for developing a green heterogeneous network}
{Islam \headeretal: Energy-cost aware off-grid base stations with IoT devices for developing a green heterogeneous network}

\corresp{Corresponding author: Md. Sanwar Hossain (e-mail: sanwareee@gmail.com)}

\begin{abstract}
Heterogeneous network (HetNet) is a specified cellular platform to tackle the rapidly growing anticipated data traffic. From communications perspective, data loads can be mapped to energy loads that are generally placed on the operator networks. Meanwhile, renewable energy aided networks offer to curtail fossil fuel consumption, so to reduce environmental pollution. This paper proposes a renewable energy based power supply architecture for off-grid HetNet using a novel energy sharing model. Solar photovoltaic (PV) along with sufficient energy storage devices are used for each macro, micro, pico, or femto base station (BS). Additionally, biomass generator (BG) is used for macro and micro BSs. The collocated macro and micro BSs are connected through end-to-end resistive lines. A novel weighted proportional-fair resource-scheduling algorithm with sleep mechanisms is proposed for non-real time (NRT) applications by trading-off the power consumption and communication delays. Furthermore, the proposed algorithm with extended discontinuous reception (eDRX) and power saving mode (PSM) for narrowband internet of things (IoT) applications extends battery lifetime for IoT devices. HOMER optimization software is used to perform optimal system architecture, economic, and carbon footprint analyses while Monte-Carlo simulation tool is used for  evaluating the throughput and energy efficiency performances. The proposed algorithms are valid for the practical data of the rural areas. We demonstrate the proposed power supply architecture is energy-efficient, cost-effective, reliable, and eco-friendly.
\end{abstract}

\begin{keywords}
Heterogeneous network, Renewable energy, Green operator network, Weighted proportional fair. 
\end{keywords}

\titlepgskip=-15pt

\maketitle

\section{Introduction}
\label{sec:introduction}
\PARstart{T}{he} world is heading into an era of highly dense wireless network which demands specific requirements, such as, tens of Gbps data rate and around 10 years of battery life for low powered IoT devices. Moreover, enhanced energy efficiency (EE), robust connectivity, and very low latency for delay-sensitive and mission-critical applications are expected to make 5G networks very popular. The concept of sustainable development through the digital revolution has been accepted by the modern generation, which has acted as a driving force behind the surge of mobile subscriber numbers from 4.5 billion in 2013 to 7.9 billion in 2020 \cite{Ericsson2020}. It is anticipated that the next wave could well be in the field of IoT technology. As per forecasts, it is anticipated that the number of IoT connections will grow to 26.9 billion by the end of 2026~\cite{Ericsson2020}. It is estimated that by 2026, 54\% of total mobile data will be carried over 5G networks. Global total mobile data traffic has reached around 51 exabytes (EB) per month in 2020 and is predicted to rise by a factor of about 4.5 to exceed 226 EB per month in 2026 \cite{Ericsson2020}. To serve a large number of users and to run the massive volume of data traffic with diverse quality of service (QoS) requirements, mobile operators have introduced the simultaneous operations of macro, micro, pico, and femtocells, which together form heterogeneous networks (HetNet). Typical coverage area radius for macro, micro, pico and femtocells are 5-30km, 1-2km, 200m and  10m, respectively. The coverage distance varies depending on the frequency, bandwidth of the signals and physical obstructions in the area. Along with coverage, these BSs consume a proportional amount of power. Currently, nearly 1.5 million 5G and 6 million 4G base stations are in operation globally ~\cite{China,5G}. 

Cellular base stations are the main consumers of the energy used by mobile operators, e.g., around 57\% as mentioned in ~\cite{en10030392}. A recent estimate of energy consumption and greenhouse gas (GHG) emission by mobile networks are around 130 TWh and 110 $MtCO_2e$ per year, respectively. The total annual carbon footprint is about 200 $MtCO_2e$, when the mobile phone emissions are included ~\cite{Ericsson2020}, and this amount has been increasing at a yearly rate of 10\% ~\cite{GSMA}. According to ~\cite{Consumption}, 11\% of the world's energy generation source is renewable energy sources (RESs). At this rate, annually at least 12.1 $MtCO_2e$ GHG emissions have been reduced due to the use of RESs in the mobile industry. Furthermore, according to the International Energy Agency, worldwide energy demand is estimated to increase by around 9\% from 2019 to 2030, which is equivalent to 1870 million tons of oil ~\cite{Outlook}. However, the reservations of global coal, oil, and natural gas remaining, warrant the needs for alternative sources of energy  ~\cite{Data}, e.g., renewable energies. High QoS at  low costs is a challenge for cellular industry services because the pricing of cellular services has been reducing gradually ~\cite{Dutta}. The use of RES for powering the BSs can be a solution to this challenge. As such, efficient energy usage is one of the crucial foci of mobile industries aiming at the reduction of costs and emission-intensive carbon footprints while maintaining a guaranteed QoS. Therefore, bringing in renewable energy (RE) from the locally available sources has become an important aspect \cite{su12093536, 7470944, 6949027}.

To analyze and validate the performance of the proposed algorithms, data collected from a practical case scenarios at a rural area in Bangladesh are considered. Bangladesh is a developing country, located between 33$^{\circ}$ N and 39$^{\circ}$ N latitude and between 124$^{\circ}$ E and 130$^{\circ}$ E longitude, where about 63.37\% of the population live in the rural areas but only 59\% of them have access to electricity ~\cite{Sustainable, Sustainable1}. The country has immense prospects of RESs such as solar, wind, biomass (agriculture residue), and geothermal ~\cite{Sustainable, 8628093}. As a result, it can be an ideal area to propose a RES based power generation infrastructure. Bangladesh has an average sunlight intensity within the range of 4-5 kWh/$m^2$ per day as well as more than 350 prospective oil-bearing crops that harvest biodiesel, for example, J. curcas, sunflower, sesame, castor, cottonseed, and groundnut oils ~\cite{DHALL2018135}. An estimated 400 EJ/yr   capacity of biomass can be contributed to global energy supply by the year 2050 that will help in ensuring the adaptability of SDG7 (affordable and clean energy) ~\cite{DHALL2018135}. According to ~\cite{HALDER20151636}, potential biomass sources in Bangladesh are 213.81 million tons where energy content is 1344.99 PJ, possible electricity generation from the source is 373.71 TWh which is equivalent to the consumption of 45.91 million tons coal or 34.01 billion cubic meters of gas. With the proper modeling and help from modern technologies, Bangladesh can harvest around 70 PWh and 7682 GWh electrical energy per year from the solar and biomass resources respectively by 2030 ~\cite{ALAMHOSSAINMONDAL20111869, HUDA2014504}. Therefore by integrating solar and biomass energy sources with adequate energy storage devices can be used to ensure a reliable RE-based supply system. Overall low electrification progress in rural areas of developing and underdeveloped countries and locally available RESs take the attention about proposing an off-grid RE aided base stations architecture.

At present, cellular network offers a variety of applications that are reshaping the digital world. Such applications include distance learning, secure home, autonomous transportation, augmented reality, and virtual reality. Moreover, massive internet of things (IoT) applications in industry, agriculture, healthcare, education, traffic, finance, environment are the near future applications of 5G. These applications have different QoS requirements, for example, some are delay-sensitive, energy-sensitive, real-time (RT) applications, and non real-time (NRT) applications. Large-scale deployment and supervision of IoT devices are becoming a crucial topic in the present day. To facilitate internet connectivity for the enormous amount of low throughput devices,  3rd Generation Partnership Project (3GPP) has proposed a new radio technology standard well-known as narrowband IoT (NB-IoT). This NB-IoT specification can implement power saving mode (PSM) and extended discontinuous reception (eDRX) mechanisms to attain a long battery life. This eDRX is an extended form of discontinuous reception (DRX) operation. DRX operation was included in 4G to achieve power saving and prolonged battery life. With intelligent and optimum use of eDRX, PSM, and DRX mechanisms, a resource scheduling algorithm can be proposed for IoT and NRT applications with an optimal tradeoff between latency and power saving.

The focus of this work is to develop a hybrid solar PV/BG focused off-grid energy sharing based supply system for powering the LTE HETNET considering the dynamic profile of RESs and traffic intensity. However, because empirical data about the 5G network is unavailable, we used LTE HetNet data for the experiments.  However, the proposed system can directly be applied to 5G data. To validate the model a rural area named Saldanga Union Parishad of Debiganj Upazila under the  district Panchagarh in Bangladesh  has been selected, which is located at  $26^\circ 7.1'0''$ North, $88^\circ 45.6' 0''$ East. This model can be equally valid for similar areas across the world depending on the available resources. As far as the best knowledge of authors, this is the first attempt to develop  model with the energy efficiency, cost, reliability, and carbon footprint analysis integrating eDRX, PSM, and DRX mechanisms and off-grid energy cooperation policies for the heterogeneous green wireless networks. 

The key contributions of this study are summarized below:
\begin{itemize} 
	\item A novel off-grid hybrid solar PV/BG system is proposed for operating an LTE HetNet in a rural area with reliable services. 
	\item An energy sharing model through an external low loss resistive power line is proposed for the HetNet cellular environment that ensures the reliability and efficient utilization of RESs. 
	\item A resource scheduling algorithm is proposed for the NRT and specifically IoT applications that achieves a large amount of power-savings through permissible delay in comparison with the existing model. 
	\item HOMER optimization software is used to obtain an optimal system architecture of the proposed framework and to perform the energy yield analysis, cost analysis and carbon footprint analysis. 
	\item A Monte-Carlo based simulation is used to evaluate the model in terms of throughput and energy efficiency under different bandwidths (BWs) allowing the possibility of dynamic traffic demand, inter-cell interference, and shadow fading.  
\end{itemize}

This work is a combined proposal of energy efficient techniques from the perspective of both energy generation and consumption sides. From the generation point of view, we propose PV/BG system architecture for operating a rural LTE HetNet with a reliable and novel energy sharing policy. From the users consumption point of view, a sleep mode based resource scheduling algorithm has been proposed for both NRT and IoT applications. The algorithm proposes a DRX mechanism, which achieves better power savings compared to existing mechanisms (as the simulation results show). Overall, to the best our knowledge, it is a novel approach for efficient energy usage for both generation (network) and consumption (user equipment) sides.

The rest of the paper is organized as follows: Section~\ref{sec2} gives a brief review of related works. Section~\ref{sec3} presents the system architecture and mathematical models of the proposed HetNet architecture. Section~\ref{sec4} summarises the system implementation and cost modeling. Section~\ref{sec5} analyses the results with the simulation setup. Finally, Section ~\ref{sec6} concludes the paper by summarizing the main outcomes.

\section{Literature Reviews}\label{sec2}
The use of renewable energy efficiently in the field of mobile cellular communication is one of the trending research topics. Moreover, research on large scale deployment of energy-efficient IoT devices will highly influence the future mobile industry. 
A standalone solar PV with sufficient storage device powered cellular BS has been introduced in ~\cite{en10030392, 7470944, 8289051}, considering real-time dynamic solar radiation profile. Although in these studies energy efficiency (EE) is critically justified with both technical criteria and financial viability, in the issue of reliability it has some drawbacks. Therefore, authors in ~\cite{8731948} and ~\cite{Alsharif2015} have suggested the incorporation of diesel generator (DG) as a non-renewable energy source with solar PV for fueling the off-grid cellular BSs. But this system increases the carbon footprint. Moreover, transportation of fuel for DG in remote areas is  not economically efficient. An integration of multiple RESs has been proposed in ~\cite{Hossain2020,su12229340, 8482262}. In~\cite{Hossain2020}, a solar PV and BG based hybrid supply model has been proposed for an off-grid homogeneous cellular network. In ~\cite{su12229340}, proposed a hybrid solar PV/wind turbine (WT)/BG powered energy supply system has been proposed for macro BSs in Bangladesh. The works in ~\cite{8482262} and \cite{en11010099} use a solar PV and WT coupled supply system framework with and without an energy sharing policy for a homogeneous BSs setup. It is clear from ~\cite{ALGHUSSAIN2020102059}, an integration of hybrid RESs with storage devices can assure reliability and minimize the gap between consumption and generation of energy. For example, in an urban area in Germany, Nokia Siemens has established a mobile BS setup powered by hybrid solar PV/WT~\cite{Nokia}.

\begin{figure*}[htb]
	\centering
	\includegraphics[width=1.5\columnwidth,keepaspectratio]{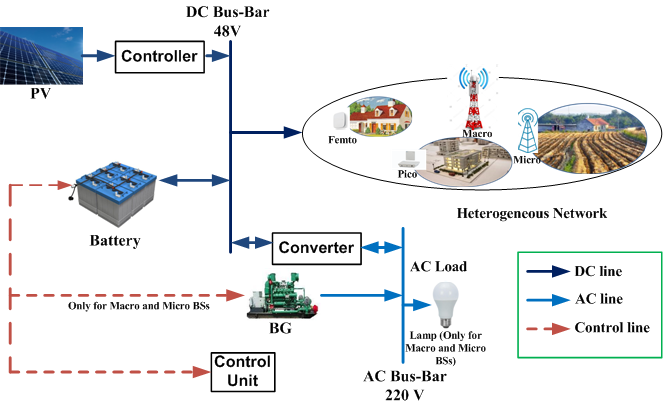}\\
	\caption{A schematic diagram of the proposed system.}\label{Fig2}
\end{figure*}
 
 Energy sharing between the adjacent cells through resistive power lines or smart grid has been used in ~\cite{7779131, 6874568, 7860175}. Agglomerative and divisive hierarchical clustering algorithms are established with average energy affinity (AEA) and stochastic energy affinity (SEA) metrics to propose an energy sharing model among BSs in ~\cite{7779131}. A RE-based community scale energy planning has been proposed by ~\cite{UGWOKE2021102750} to get a reasonable  reduction in energy use and GHG emissions. In~\cite{AWAD2018221}, the authors proposed an optimized community shared solar PV framework. An optimal energy cooperation framework among cellular BSs has been proposed in ~\cite{6874568}, where both RE and grid have been incorporated using physically connected resistive power lines. An energy cooperation model between macro BSs has been demonstrated in ~\cite{7860175}. From the aforementioned research articles, it is clear that on-site green energy sharing between BS's is affordable and it makes the systems reliable and cost-effective.

An efficient resource scheduling algorithm can optimize the tradeoff between different QoS requirements for specific applications. A minimum guaranteed throughput scheduling  algorithm for NRT applications has been introduced in ~\cite{4392070}. One of the most popular solutions for resource scheduling of NRT applications is the proportional fair (PF) scheduler which successfully incorporates cell throughput and fairness. However, it ignores the queue size and does not attempt to improve power saving from DRX operation. A channel-adapted and buffer aware (CABA) packet scheduling algorithm has been proposed ~\cite{4677944}, which considers the queue size. But this stand-alone proposal cannot take all the advantages of the popular PF scheduler. The work in ~\cite{6671166} considers energy metrics at the base station, with a certain rate guarantee for each mobile station. A radio resource allocation framework is proposed in ~\cite{6503989} to ensure proportional fairness among the user equipment (UE). A few solutions that make such resource allocators capable of supporting both guaranteed bit-rate and best-effort traffic are proposed in ~\cite{7194064}. A new scheduling scheme named proportional equal throughput (PET)~\cite{7034111}, which offers better fairness among users without reducing average user throughput, in comparison with other scheduling algorithms. 

In ~\cite {REDDY2020102428}, a context-conscious sleep and wake-up cycle of the fog nodes have been proposed in the target of energy minimization at the fog layer. Authors in ~\cite{4657144, 6719041} analyzed the effects of the DRX parameter on UE's energy saving and latency. Nokia has an exploration on industrial DRX, which is illustrated in ~\cite{6363708}. It is well recognized that power-saving and delay are always a tradeoff. To optimize the tradeoff, several DRX algorithms have been introduced in ~\cite{6399035, 6554626, 6839948, 6921789, 6566901}. On the other hand, PSM and eDRX are two solutions to optimize the device power consumption proposed for EC-GSM, LTE-M, and NB-IoT technologies for delay insensitive but energy sensitive applications which have been introduced in 3GPP Rel.12. In ~\cite{8120238}, the authors have done an energy consumption analysis over an NB-IoT device under different traffic intensities and coverage areas. However, to the best of our knowledge, there is no such algorithm proposed, where both PSM and eDRX have jointly analyzed to ensure optimum battery lifetime. 

\section{System Architecture and Mathematical Modeling}\label{sec3}

\begin{figure*}[t]
	\centering
	\includegraphics[width=1.5\columnwidth,keepaspectratio]{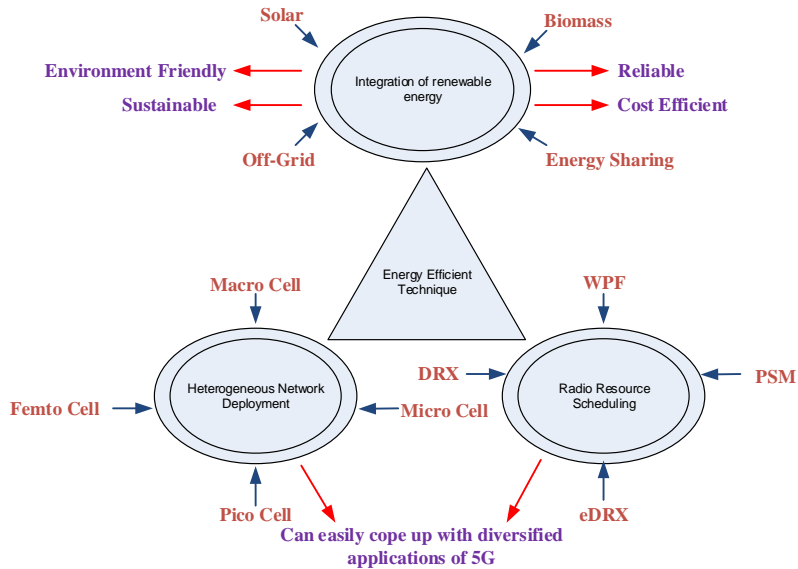}\\
	\vspace{-7.0cm}
	\caption{ An overview of the energy-efficient techniques of the proposed model.}
	\label{Fig1}
\end{figure*}

\begin{table}[hbt]
	\centering
	\caption{A list of the notations.}
	\tiny
	\label{notations}

	{
	\begin{tabular}{p{2cm} p{4cm}}
	\toprule
	\textbf{Symbol} & \textbf{Meaning} \\
	\toprule

    $CC$ &	Capital cost \\
    $Comp$ &	Lifetime of the component \\
    
	$CRF$ &	Capital recovery factor \\
	$FC$ &	Fuel cost \\
	
	$NPC$ &	Net present cost \\
	$OMC$ &	Operation and maintence cost \\
	
	$PSH$ & Peak solar hour \\
	$RC$ &	Replacement cost \\
	
	$Rem$ &	Remaining lifetime \\
	$SINR$ & Signal-to-interference plus-noise-ratio \\
	
	$SV$ &	Salvage value \\
	$TAC$ &	Total annualized cost \\
	
		\bottomrule
\end{tabular}
}
\end{table}

\begin{table*}[hbt]
	\centering
	\caption{A list of the symbols.}
	\tiny
	\label{Summary}

	{
	\begin{tabular}{p{1cm} p{4cm}|| p{1cm} p{4cm}}
	\toprule
	\textbf{Symbol} & \textbf{Meaning} & \textbf{Symbol} & \textbf{Meaning}\\
	\toprule
	
    $\delta_p$	& Load dependency power gradient &
    $\eta_{BM}$	& Efficiency of the biomass generator \\
    
    $\eta_{EE}$ & Energy efficiency &
    $\eta_{eDRX}$ & Number of paging intervals in total eDRX period \\
    
    $\eta_{PV}$	& Panel efficiency &
    $\sigma_{cool}$	& Loss component of cooling system \\
    
	$\sigma_{DC}$	& Loss component of DC-DC regulator &
	$\sigma_{MS}$	& Loss component of main supply \\
	
	$\tau_D$ & Data transfer timer &
	$\tau_{DRX}$ & DRX timer \\
	
	$\tau_{DRX}^\prime$ & End of DRX timer &
	$\tau_{eDRX}$ & eDRX timer \\
	
	$\tau_{eDRX}^\prime$ & End of eDRX timer &
	$\tau_{i}$ & Radio Resource Control (RRC) inactivity timer \\
	
	$\tau_{i}^\prime$ & End of inactivity timer &
	$\tau_L$ & eDRX long cycle timer \\
	
	$\tau_{PSM}$ & PSM timer &
	$\tau_{PSM}^\prime$ & End of PSM timer \\
	
	$\tau_S$ & eDRX short cycle timer &
	$\tau_{st}$ & $\frac{\tau_L-\tau_S}{\eta_{DRX}}$ \\
	
	$\chi$ & Poisson packet in queue &
	$\chi_{i_{UL/DL}}$ & UL/DL poisson packet in queue \\
	
	$\chi_{i_{UL}}$ & UL poisson packet in queue &
	$B_{aut}$ &	Battery autonomy \\
	
	$B_c$ &	Capacity of the battery bank &
	$B_{DOD}$	& Depth of discharge of the battery bank \\	
	
	$B_{SOCmin}$ & Lower threshold limit of battery discharge &
	$C_f$ &	Capacity factor \\
	
	$CV_{BM}$	& Calorific value of the biomass &
	$D$	& Backup days \\
	
	$E_{BS}$ &	Annual BS load consumption &
	$E_{BG}$ &	Renewable energy generated from BG \\
	
	$E_{BG}$ & Generated electricity from BG &
	$E_{batt}$ &	Energy afforded by the battery bank \\
	
	$E_{battmax}$ &	Maximum capacity of the energy storage  devices &
	$E_{battmin}$ &	Minimum capacity of the energy storage devices \\
	
	$E_{CS}$ &	Yearly capacity storage &
	$E_D$ & Total demand of electricity \\
	
	$E_{ED}$ & Yearly energy deficiency &
	$E_{excess}$ & Excess electricity for critical time use (at least 10\% of  $E_{batt}$)\\
	
	$E_{gen}$ &	Total generated electricity &
	$E_l$ & Sum of converter, battery bank and energy transfer losses (kWh/yr) \\
	
	$E_{pv}$	& Energy generated from solar PV array (kWh) &
	$E_{Share}$	& Shared energy (Permissible up to 90\% of Ebatt) \\
	
	$I$ & Current through the conductor &
   	$K_b$ &	Battery capacity co-efficient factor \\
   	
   	$L_{BS}$ &	Average daily BS load in kWh &
   	$L_{batt}$	& Lifetime of the battery bank \\
   	
   	$N$ & Total number of BSs &
   	$N_{batt}$ &	Number of batteries in the battery bank \\
   	
   	$P_{BG}$	& Annual generated power &
    $P_{BS}$	& BS power consumption \\
    
    $P_0$ &	Power consumption at idle state &
	$P_{PA}$ & Power amplifier power consumption (kWh/yr) \\
	
	$ P_{RF}$ & RF power consumption &
	$P_{sleep}$	& Power consumption at sleep mode \\
	
	$P_{TX}$ & Maximum transmission power (W) &
    $ P_{BB}$ & Baseband power consumption \\
    
    $Q_{nom}$ &	Nominal capacity of a single battery (Ah) &
    $R_{batt}$	& Battery float life (year) \\
	
	$R_{total}$ & Total achievable throughput &
	$R(l)$ & Resistance of the l km conductor length \\ 
    
    $R_{pv}$	& Rated Capacity for solar PV panel (KW) &
    $T_a$	& Annual battery throughput (kWh/year) \\
	
    $T_{batt}$ &	Lifetime throughput of a single battery (kWh) &
    $T_{BM}$	& Average biomass availability (t/year) \\
    
    $t_{op}$ &	BG running hours &
    $U$	& Number of user equipments \\
    
	$V_{nom}$ &	Nominal voltage of a single battery (V) &
	$V$	& DC bus-bar voltage (Volt) \\
	\bottomrule
\end{tabular}
}
\end{table*}

A schematic diagram of the proposed hybrid solar PV/BG powered rural area-based off-grid HetNet is illustrated in Figure ~\ref{Fig2}. The HetNet consists of macro, micro, pico, femto type BSs, where each type of BS has its specific size, power consumption, and data rate requirements according to the required QoS. Generally, a small size BS requires small radiated power due to its small coverage area. Hence, pico and femto BSs require low power consumption, they are supplied with solar PV alone. In contrast, macro and micro BSs require high power consumption. As such they are connected to both solar PV and BG with BG only activated if PV cannot supply enough power. To ensure a continuous power supply, a battery bank is connected to overcome the shortage or outage of RES. Here, the power supply to each base station is controlled via an energy management unit (EMU). The EMU is also connecting renewable energy sources and energy storage devices. EMU primarily collects the generated power from renewable energy sources and distributes the generated energy according to the respective base station requirements. EMU also avoids overvoltage and overcurrent. It consequently improves battery life by minimizing intense discharge from the storage unit. A BS is a DC load. However, there are some AC load such as a lamp and air cooler that need to be powered. Converters are used to convert DC into AC and vice versa as needed. This section presents the proposed system model along with the renewable energy generation model, BS power consumption model, and energy sharing algorithm in the context of off-grid LTE cellular networks. Notations and symbols that are used in the subsequent sections are summarised in ~Tables \ref{notations} and \ref{Summary} respectively.

An overview of the energy-efficient techniques from the prospective of both energy generation and consumption are shown in Figure ~\ref{Fig1}. We illustrate an overall architecture of the proposed system and methods to establish our proposed energy-efficient techniques from the perspective of both energy generation and consumption sides. By integrating renewable energy sources and HetNet deployment, we make the system energy-efficient from the perspective of power generation. On the other hand, by radio resource scheduling, we optimize the end device energy consumption. Overall, in order to consider these three aspects, the energy-efficient technique has been shown using a triangle sign.

\subsection{Solar PV panel}

\begin{figure}[tb]
	\centering
	\includegraphics[width=0.9\columnwidth, keepaspectratio]{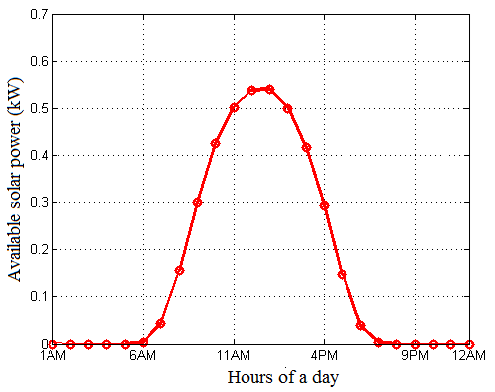}\\
	\caption{Average hourly solar power production over a day for 1 kW PV module size.}			\label{solar}
\end{figure}

Solar PV produces electricity from solar energy directly. The amount of solar energy generation heavily depends on the geographic location (availability of direct sunlight), beam radiation, panel materials, and tracking mode. On focusing the specifications and capital cost ‘Sharp ND-250QCs (poly crystalline)' is selected in this work. Additional information of the Sharp ND-250QCs (poly crystalline) solar module can be found in ~\cite{Sharp}. The temporal variation of solar power generation is calculated in Figure~\ref{solar} using the System Advisory Model (SAM) for 1 kW PV module size~\cite{SAM}.

\subsection{Biomass Generator}

The amount of energy generated by the BG mostly depends on the average biomass availability, the calorific value of the biomass, the efficiency of the BG, and BG running hours. The combination of BG with the solar PV panel plays an important role in compensating the BS energy demand during the malfunction of RES.

\subsection{Battery Bank}

\begin{figure*}[htb]
	\centering
	\includegraphics[width=1.7\columnwidth]{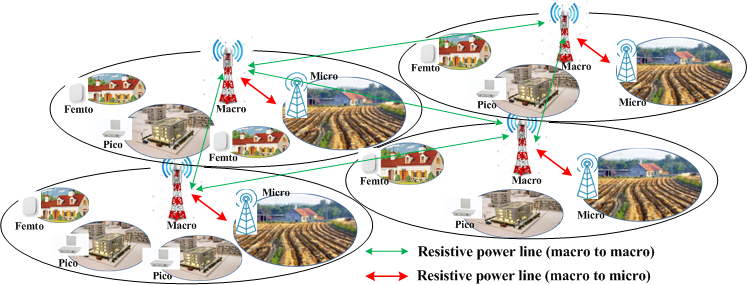}\\
	\vspace{0cm}
	\caption{Energy-sharing model among the adjacent BSs of HetNet.}			\label{model}
\end{figure*}

Battery storage is considered  a crucial component in the  sustainable energy. 
The battery bank autonomy ($B_{aut}$) specifies the credible time limit up to that the storage setup can deliver the necessary electricity to run BS as load if the RES is unavailable. It can be calculated as follows \cite{en11061500, su12229340}:

\begin{equation}
B_{aut} = \frac{N_{batt} \times V_{nom} \times Q_{nom} \times (1 - \frac{B_{SOC_{min}}}{100}) \times (24~\text{h/day})}{L_{BS}}	,\label{Eq:BAUT}
\end{equation}

The capacity of the battery bank ($B_c$) can be expressed as follows \cite{en11061500, su12229340}: 

\begin{equation}
B_c=\frac{P_{BS} \times D \times t}{B_{DOD} \times V \times K_b} Ah ,\label{Eq:BC}
\end{equation}

\subsection{Reliability analysis}

Annual capacity storage ($E_{CS}$) or ‘annual capacity shortage’ is a variable used to quantify the system's reliability. It can be calculated as:

\begin{equation}
{E_{CS}}= \frac {E_{ED}}{E_{BS}} 
\end{equation}

${E_{ED}}$ can be expressed as follows and calculated in kWh/yr:

\begin{equation}
{E_{ED}}={E_{BS}}-{E_{gen}}
\end{equation}

where ${E_{gen}}$ is the generated electricity, which can be written for macro and micro BSs:
\begin{equation}
{E_{gen}}={E_{PV}}+{E_{BG}}  
\end{equation}

For pico and femto BSs the above equation can be modified as follows:
\begin{equation}
{E_{gen}}={E_{PV}}  
\end{equation}

To ensure  reliability of the system, the proposed scheme is designed with sufficient backup power reserve over the entire project duration. The excess electricity, generated as backup power, can be estimated by:

\begin{equation}
{E_{excess}}={E_{gen}}-{E_{BS}}-{C_{loss}}-{B_{loss}}  
\end{equation}

where, $C_{loss}$ and $B_{loss}$ represent the losses associated with the converter and battery respectively. 

\subsection{Energy-Sharing Model}

\begin{figure*}[htb]
	\centering
	\includegraphics[width=1.8\columnwidth]{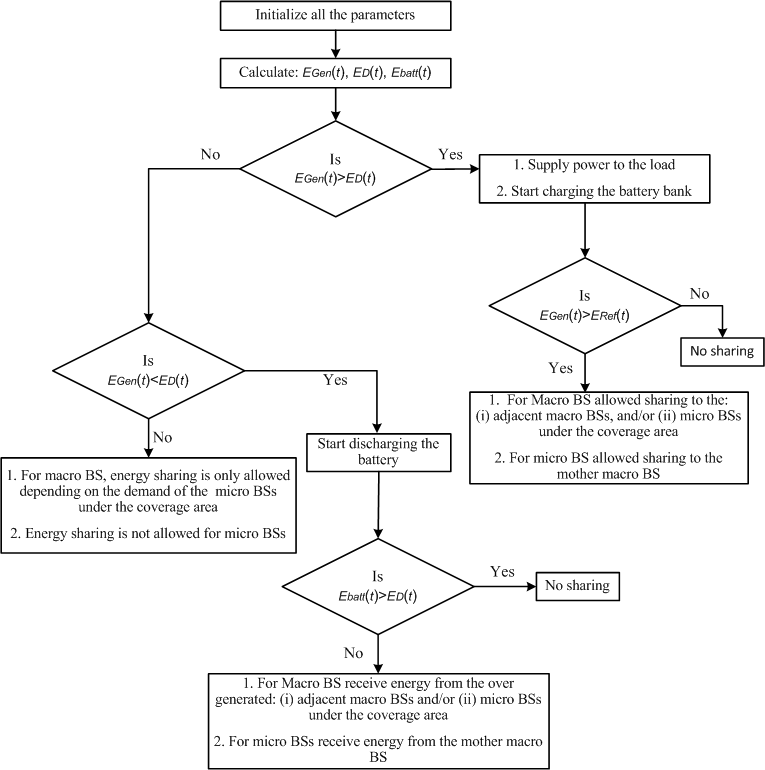}\\
	\caption{Flow diagram of the proposed energy sharing model.}	\label{Energy_sharing}
\end{figure*}

This section proposes an energy-sharing algorithm to ensure green, optimal, and uninterrupted power supply for an off-grid HetNet configuration. The basic features of the macro, micro, pico, and femto cells are presented in Table \ref{Het}.


\begin{table}[htb]
	\centering
	\caption{Cell size and transmitted power for a heterogeneous network \cite{Auer1}.}
	\label{Het}
	{
		\begin{tabular}{l l l}
			\toprule
			\textbf{Cell type} & \textbf{Cell size} & {\textbf{Power consumption}} \\
			\midrule
			Macro & 1-30 km & Tens of Watts \\
		    Micro & 0.4-2 km & 1-6.3 W\\
			Pico & 4-200 m & 200 mW-2 W\\
			Femto & 10 m & 20-200 mW\\
			\bottomrule	
	\end{tabular}}
\end{table}

To propose the algorithm, an energy sharing model is used, which is illustrated in Figure \ref{model}. In the core of HetNet there are cells. A cell is formed with a macro base station (BS) placed at the centre and one or more micro BSs are connected to the macro BS in a star topology. There can be multiple micro BSs inside the cell of a macro BS. Because of the star connection arrangement, each of these smaller BSs is directly connected to the central macro BS. As such, a macro BS can share energy with all other micro BSs in the cell. A macro BS can also share energy with other adjacent macro BSs directly connected in the network. All the connections for energy sharing between the BSs has been intended to do through low resistive power lines and in every case it is half-duplex. In the HetNet, different types of cells are required for various applications with various quality of service (QoS) requirements. The distance between the cells is small in the heterogeneous network. As a result, the low resistive power line method can be practicable for sharing energy \cite{han2020energy}. This connection for energy sharing may effectively use BSs' renewable energy sources. 

Through the energy sharing policy, the excess electricity will be shared according to the flow diagram illustrated in Figure \ref{Energy_sharing}. Here, $E_{ref}$ is the amount of excess electricity that is 10\% greater than the load requirement. The resistive value of the proposed connection has been taken from the American Wire Gauge (AWG) standard conductor size table \cite{Solaris}.

The total shared energy ($E_{Share}$) to the adjacent BSs can be calculated as \cite{8731948}:

\begin{equation}
E_{Share} =E_{excess}-E_{loss},   \label{Eq:Eshare1}
\end{equation}

where $E_{loss}$ is the path loss because of the resistance of the conductor, can be calculated as \cite{8731948}:

\begin{equation}
E_{loss} = I^{2} R(l)\times t_{r} = \frac{P_{Share}^2 R(l)}{V^{2}} \times t_{r} \label{Eq:ELINE}
\end{equation}

The percentage of energy savings because of  energy sharing policy can be calculated as \cite{8731948}:

\begin{equation}
 E_{Save}(\%) = \frac{\sum_{i=1}^{N} E_{Share}(t)}{\sum_{i=1}^{N} E_{BS}(t)} \times 100\% \\ 
\label{Eq:Esav}
\end{equation}

\subsection{Throughput and Energy Efficiency Model}	

The performance of the wireless network is evaluated in terms of throughput and EE metrics. According to Shanon’s information capacity theorem, total achievable throughput in a network at time $t$ can be expressed by \cite{8731948}:

\begin{equation} 
R_{total}(t) = \sum_{k=1}^{U}\sum_{i=1}^{N} BW log_2(1+SINR_{i,k}) \label{Eq:THP}
\end{equation}

EE is defined as the ratio of total network throughput and total power consumed by the network. Thus, the EE metric denoted as $\eta_{EE}$ for time $t$ is defined as \cite{8731948}:

\begin{equation}
\eta_{EE}= \frac{R_{total} (t)}{P_{BS} (t)}  \label{Eq:EE}
\end{equation}

Here, the received signal to interference plus noise ratio (SINR) at $k^{th}$ user equipment from $i^{th}$ BS can be given by
 
 \begin{equation}
SINR_{i,k} = \frac{P_{rx}^{i,k}}{P_{k,inter}+P_{k,intra}+P_N} \\ 
\label{SINR}
\end{equation}			

where, $P_{k,intra}$ is the intra-cell interference, $P_{k,inter}$ is the inter-cell interference is the transmitted power in dBm,   $P_N$ is the additive white Gaussian noise (AWGN) power given by $P_N=-174+10log_{10}(BW)$  in dBm with BW is the bandwidth in Hz. However, the orthogonal frequency division multiple access (OFDMA) technique in the LTE system results in zero intra-cell interference.

\subsection{Mathematical model of costs}
Net present cost (NPC) is the total cost of the proposed system during the full life cycle, which can be calculated as follows \cite{en11061500}:   

\begin{equation} \label{Eq:NPC}
NPC = \frac{TAC}{CRF} 
= CC + RC + OMC + FC -SV,
\end{equation}

whereas total annualized cost (TAC) and capital recovery factor (CRF) can be expressed by equations (\ref{Eq:TAC}) and (\ref{Eq:CRF}) respectively \cite{en11061500}.

\begin{equation} \label{Eq:TAC}
TAC = TAC_{CC} + TAC_{RC} + TAC_{OMC}
\end{equation}

\begin{equation} \label{Eq:CRF}
CRF = \frac{i (1+i)^{L}}{(1+i)^{L}-1}
\end{equation}

Here, $L$ expresses the lifespan of the proposed project and $i$ expresses the every twelve months interest rate. Instead, the salvage value, only applicable for those components which have a longer lifespan than project lifespan, can be expressed by \cite{en11061500},

\begin{equation}
SV = Rep (\frac{Rem}{Comp}),   \label{Eq:SV}
\end{equation}

\subsection{Problem formulation and optimization}
The proposed hybrid energy generation model is considered as an optimization problem, where the target is to minimize NPC by ensuring maximum utilization of RESs. The number of solar PV, biomass generator, battery bank, and converter is optimized according to the operational constraints and load demands. Specifically, each watt of generated energy should be utilized. For an effective performance of the optimization process to formulate the optimal hybrid power system, a number of parameters for system components, such as operational lifecycle, component efficiency, and associated cost, are taken into account. At each hour, using HOMER optimization software we determine how to satisfy electricity demand using HetNet infrastructure, including losses, as well as provide backup power. It also calculates the BS load demand to the input power supply in each iteration, and the optimization phase follows all simulations. Finally, we identify the least expensive component combination that meets the BS load while ensuring no power shortage. The following optimization process was maintained during the simulation process

\textbf{For macro and micro BSs:} 
\begin{mini!} 
	{}{NPC}  
	{}{}{}{}{}{}
	\addConstraint{E_{PV} + E_{BG} > E_{BS}} 
	\addConstraint{E_{PV} + E_{BG} + E_{batt} = E_{BS}+ E_{l}}
	\addConstraint{E_{Share} = E_{Gen}+ E_{batt}- E_{BS}- E_{l}-E_{Excess}}
	\addConstraint{E_{battmin} \leq E_{batt} \leq E_{battmax}},	
\end{mini!}

\textbf{For pico and femto BSs:}
\begin{mini!} 
	{}{NPC}  
	{}{}{}{}{}{}
	\addConstraint{E_{PV} > E_{BS}} 
	\addConstraint{E_{PV} + E_{batt} = E_{BS}+ E_{l}}
	\addConstraint{E_{battmin} \leq E_{batt} \leq E_{battmax}},	
\end{mini!}

where $E_l$ is the sum of converter losses, battery bank losses, and surplus energy transfer losses in kWh/yr. For macro and microcell configuration, the hybrid solar PV and BG can support the BS energy demand as mentioned in (18b). In the case of pico and femtocell configuration, the surplus energy transfer losses is zero and the standalone solar PV system can fulfill the BS entire energy requirement as mentioned in (19b). The constraint in (18c) and (19c) ensures that the annual energy obtained by the renewable energy sources carries the annual BS energy consumption with associated losses. The amount of surplus electricity is preserved for future use during the abnormal condition or shared among neighboring BS due to scarcity of energy is described by the constraint (18d). Constraint (18e) and (19d) indicates the battery bank storage capacity should not exceed the maximum limit and not reached the below threshold level.

\subsection{Proposed resource scheduling algorithm}	

A resource scheduling algorithm attempts to make an appropriate apportionment of the resources with fulfilling application-wise QoS requirements. Some general requirements are optimized spectral efficiency ensuring cell throughput, fairness, minimum interference for cell-edge users, and optimum load balancing. PF is one of the well-known schedulers, especially for NRT applications. It deals with two competing qualities, not only maximizes the total throughput of the network but also ensures at least a minimal level of service. But better latency, user energy efficiency, and application wise priority cannot be assured by PF. LTE establishes separate resource blocks (RBs) for RT and NRT applications. The RT application includes voice, live streaming video, and online gaming. The NRT application includes buffered streaming video, web browsing, e-mail, chat, FTP, and P2P file sharing. The BSs are fully conscious of the bearers. 

In our proposed model, the available resources are divided into two parts for the use of RT and NRT services depending on the data rate demand of the users. For RT applications, resources can be scheduled using the existing VT-M-LWDF scheduler, M-LWDF scheduler, or any other popular solution that considers the packet delay. In the proposed model, the main focus has been given on NRT applications. A weighted proportional fair (WPF) scheduling algorithm has been proposed for NRT applications, where the DRX mechanism has been applied, weight will be determined based on the applications and queue size. 

\begin{equation}
\text{WPF metric} =  \text{WF} \times \text{PF metric} 
\end{equation}
\vspace{-3mm}
\begin{equation}
\text{WF} = A_k(t)\times Q_k(t)\times T_y^{Inac} 
\end{equation}

The amount of pending data or queue size $Q_k(t)$ can be used to indicate the data rate requirement. $A_k(t)$ represents the priority of user applications. $T_y^{Inac}$ is the inactivity timer that represents the way of power-saving for resource allocation.  In the case of downlink (DL), the queue size $Q_k(t)$ is already known to BS. On the other side, the UE sends pending data status or buffer status report (BSR) to the BS in case of uplink (UL). BSR index represents the buffer content in between a minimum and a maximum range. In our proposed model, $Q_k(t)$ is the average of the maximum and minimum buffer size as follows:

\begin{equation}
Q_k(t) = \frac{BS_k^{min}(t)+BS_k^{max}(t)}{2} 
\end{equation}

\begin{algorithm*}[htb]
	\centering
	\caption{: Algorithm for the communication between $i^{th}$ NRT application based device and BS.}
	\label{Algorithm1}
	{\tabulinesep=1.2mm
		\begin{tabu}{l}
			1: \hspace{1mm} Initialize: $\tau_D$, $\tau_S$, $\tau_L$, $\tau_{st}$, $\eta_{eDRX}$, $\tau_{i}$, $\tau_{i}^\prime$, $\tau_{DRX}$, $\tau_{DRX}^\prime$, $\chi_{i_{UL/DL}}$, and $\chi_{i_{UL}}$\\ 
			2: \hspace{1mm} \textbf {for} $i=1:k$   \hspace{1cm} //$k$ is the number of total users\\
			3: \hspace{7mm} \textbf {while} $\chi_{i_{UL/DL}}>0$ \hspace{5mm} //data transfer request is present between $i$ user and BS\\
			4: \hspace{12mm} Start data transfer ($\tau_D$) \\
			5: \hspace{12mm} update $\chi_{i_{UL/DL}}$  \\
			6: \hspace{7mm} enable inactivity timer ($\tau_{i}$)  \\
			7: \hspace{7mm} \textbf {if}  $\chi_{i_{UL/DL}}>0$ \&\& $\tau_{i}<\tau_{i}^\prime$  \hspace{0.5cm} //more data to transmit within the inactivity timer\\
			8:\hspace{12mm} Go to Step 3 \\
			9:\hspace{7mm} \textbf {else} \\
			10:\hspace{12mm} enable $\tau_{DRX}$ \\
			11:\hspace{7mm} \textbf {for}  $l= \tau_{DRX}:\tau_L-\tau_S:\tau_{DRX}^\prime$ \\
			12:\hspace{12mm} \textbf {for}  $m=\tau_S:\tau_L$ \\
			13: \hspace{17mm} \textbf {if}  $\chi_{i_{UL/DL}}>0$ \&\& $\tau_{DRX}<\tau_{DRX}^\prime$ \\
			14:\hspace{22mm} Go to Step 3 \\
			15:\hspace{17mm} \textbf {else} \\
			16:\hspace{22mm} $m=m+\tau_{st}$ \\
			17:\hspace{12mm} \textbf {for}  $n=\tau_L:\tau_S$ \\
			18: \hspace{17mm} \textbf {if}  $\chi_{i_{UL/DL}}>0$ \&\& $\tau_{DRX}<\tau_{DRX}^\prime$ \\
			19:\hspace{22mm} Go to Step 3 \\
			20:\hspace{17mm} \textbf {else} \\
			21:\hspace{22mm} $n=n+\tau_{st}$ \\
			22: \hspace{7mm} $i=i+1$  \\
			
	\end{tabu}}
\end{algorithm*} 

The DRX mechanism is incorporated in LTE to attain power-saving and prolonged battery life of the UE. Algorithm \ref{Algorithm1} is proposed only for NRT applications so that significant improvement in power saving attained allowing limited latency. In the algorithm, a novel DRX model has been proposed. NRT applications can support more delay than RT applications. According to \cite{3GPP}, permissible delay for NRT applications after re-transmission delay is 107ms.

UE has to be in the ‘ON’ state all the time even if there is no traffic. It is because UE has to keep listening or ready to decode PDCCH. But continuous ON state increases battery consumption drastically. DRX is a solution to address of this problem. Algorithm \ref{Algorithm1} describes the proposed communication procedure between BS and NRT applications. In RRC\_CONNECTED state, after the end of data transfer, an inactivity timer will start. Inactivity timer will reset due to DL or UL packet arrival before it expires. If the timer can run-up to its expiry with no data transfer request, the DRX cycle will start. In the existing DRX model, the UE starts DRX with some short cycles ($\tau_S$) which means some short sleep time. If the short cycles can complete successfully without any DL packet arrival at the time of paging or UL packet arrival between this period, the long cycle ($\tau_L$) will start. During the long cycle, between paging, long sleep time will continue. In the proposed model, the DRX cycle sleep time keeps growing from a short cycle timer ($\tau_S$) to a long cycle timer ($\tau_L$) by a fixed step size $\tau_{st}$. If, in total $n$ cycles are needed to reach $\tau_L$, then these $n$ cycles will also need to decline up to $\tau_S$. If during this period no packet has arrived, the fall will be followed by a rise again in a triangular fashion.  But as soon as the DL packet has arrived in the paging period or the UL packet has arrived in the whole DRX period, immediate data transfer will start and the full cycle will rotate.
We assumed an M/G/1 queue where the packet arrival rate ($\lambda$) follows Markovian (Poisson process), and the service rate ($\mu$) follows a General distribution. Traffic intensity can be calculated from $\rho=\lambda/\mu$. To calculate the percentage of power saving we have used the following equation, which has modified from \cite{6151867}

\begin{align*}
P_{Save}(\%) = \bigg[1-\bigg[\frac{E[T_{Data}]}{E[T_{Total}]}+\frac{P_{Wait}\times E[T_{Wait}]}{P_{Data}\times E[T_{Total}]} \\
+ \frac{P_{OFF}\times E[T_{OFF}]}{P_{Data}\times E[T_{Total}]}\bigg] \bigg] \times 100
\end{align*}

\begin{figure*}[hbt]
	\centering
	\includegraphics[width=1.8\columnwidth]{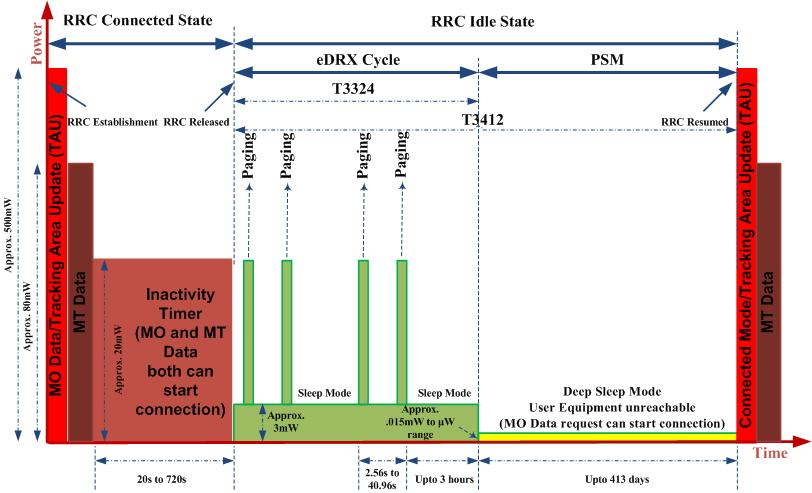}\\
	\caption{Illustration of user device’s behavior in NB-IoT (PSM and eDRX).}			
	\label{NB-IoT}
\end{figure*}

In our proposed model, an algorithm has been presented based on PSM and eDRX which are two solutions to optimize the device power consumption for NB-IoT technologies for delay insensitive but energy sensitive applications. As soon as an end device is powered on, it goes for the frequency and time synchronization with some necessary configurations from the network. This Radio Resource Control (RRC) connection setup is an important step after which the end device can exchange data with the network. When the RRC connection is established, the device is said to be in RRC Connected state. In NB-IoT, there are two RRC states for devices, namely, RRC Connected and RRC Idle as shown in Figure \ref{NB-IoT}. When the device releases its active RRC connection, it moves to the RRC Idle state. RRC connected state consists of mobile originated (MO) or mobile terminated (MT) data transfer, tracking area update (TAU), and an inactivity timer. In this state, the device consumes more energy as it gets dedicated bearers established to begin the data transmission and needs to monitor the DL channel. The expiration of the inactivity timer makes the device switch from the RRC Connected to RRC Idle state. In the RRC Idle state, NB-IoT defines two power-saving schemes, i.e., eDRX and PSM. A summary of PSM and eDRX are described below, and the details of the RRC states are illustrated in Figure \ref{NB-IoT}.

\subsubsection{PSM (Power Saving Mode)}

PSM permits UE to reduce power consumption to the most bottom level by transferring its circuitry into a deep sleep mode.  But still, the device is registered to the system. DL or MT data transfer requests can not be scheduled in this mode because the device is inaccessible for the BS. On the other hand, for MO or UL data transfer device leaves PSM mode at an instance. Moreover, it also wakes up if a request comes for a routing area update (RAU) or TAU. After every entry of PSM mode at first device runs some active timer when MT data transfer can also be scheduled by monitoring the paging channel. For NB-IoT applications, battery lifetime is a vital issue. On the other side, downlink latency does not have any significant importance for NB-IoT applications and infrequent paging is sufficient to monitor. For NB-IoT applications, the PSM cycle timer supports deep sleep up to 413 days.

\subsubsection{eDRX(extended discontinuous reception) model}

eDRX is an extended form of DRX to keep on longer in the idle or power-saving state between two paging. At first, it starts with some short active cycles and then enters into the long deep sleep cycles which minimize device power consumption. In eDRX mode, devices are periodically available for DL or MT data transfer requests. From LTE Release 13 specifications, it is defined that eDRX idle mode can be extended up to 3 hours.

In this paper, a scheduling algorithm for the communication between BS and NB-IoT application devices using PSM and eDRX is proposed in Algorithm \ref{Algorithm2}. After the end of data transfer, an inactivity timer will start. If the inactivity timer can reach its end without any data packet arrival, eDRX cycle will start. eDRX cycle consists of some short and some extended long cycles. If the eDRX cycle can run-up to the end, PSM will start. UL data transfer request of TAU during the sleep period can restart the full cycle by starting the ON or data transfer active period. DL/MT data transfer can start ON period if and only if it can reach inactivity timer or paging period. 

To calculate the percentage of power saving we have used the following equation, which has modified from \cite{6151867}.

\begin{align}
P_{Save}(\%) = \bigg[1-\bigg[\frac{E[T_{Data}]}{E[T_{Total}]}+\frac{P_{Inactive}\times E[T_{Inactive}]}{P_{Data}\times E[T_{Total}]} \\ + \frac{P_{eDRX}\times E[T_{eDRX}]}{P_{Data}\times E[T_{Total}]} + \frac{P_{PSM}\times E[T_{PSM}]}{P_{Data}\times E[T_{Total}]}\bigg] \bigg] \times 100
\end{align}

\begin{algorithm}[ht]
	\centering
	\caption{: Algorithm for the communication between $i^{th}$ IoT device and BS.}
	\label{Algorithm2}
	{
		\begin{tabu}{ll}
			1: \hspace{1mm} Initialize: $\tau_D$, $\tau_S$, $\tau_L$, $\tau_{st}$, $\eta_{eDRX}$, $\tau_{i}$, $\tau_{i}^\prime$, $\tau_{eDRX}$, $\tau_{eDRX}^\prime$,\\ $\tau_{PSM}$, $\tau_{PSM}^\prime$,  $\chi_{i_{UL/DL}}$, and $\chi_{i_{UL}}$\\
			2: \hspace{1mm} \textbf {for} $i=1:k$ \\
			3: \hspace{7mm} \textbf {while} $\chi_{i_{UL/DL}}>0$\\
			4: \hspace{12mm} Start data transfer ($\tau_D$) \\
			5: \hspace{12mm} update $\chi_{i_{UL/DL}}$  \\
			6: \hspace{7mm} enable inactivity timer ($\tau_{i}$)  \\
			7: \hspace{7mm} \textbf {if}  $\chi_{i_{UL/DL}}>0$ \&\& $\tau_{i}<\tau_{i}^\prime$ \\
		    8:\hspace{12mm} Go to Step 3 \\
			9:\hspace{7mm} \textbf {else} \\
			10:\hspace{12mm} enable $\tau_{eDRX}$ \\
			11:\hspace{7mm} \textbf {for} j=1:N \\
			12:\hspace{12mm} check $\chi_{i_{UL/DL}}$ in every $\tau_S$ interval \\
			13: \hspace{12mm} \textbf {if} $\chi_{i_{UL/DL}}>0$ \\
			14:\hspace{17mm} Go to Step 3 \\
			15:\hspace{12mm} \textbf {else} \\
			16:\hspace{17mm} j=j+1\\
			17:\hspace{7mm} check $\chi_{i_{UL/DL}}$ in every $\tau_L$ interval \\
			18: \hspace{7mm} \textbf {if} $\chi_{i_{UL/DL}}>0$ \&\& $\tau_{eDRX}>\tau_{eDRX}^\prime$\\
			19:\hspace{12mm} Go to Step 3 \\
			20: \hspace{7mm} \textbf {else} \\
			21: \hspace{12mm} Continue ($eDRX$) \\
			22:\hspace{7mm} enable Power Saving Mode ($\tau_{PSM})$  \\
			23:\hspace{7mm} \textbf {if} $\chi_{i_{UL}}>0$  \\
			24:\hspace{12mm} Go to Step 3 \\
			25: \hspace{7mm} \textbf {else} \\
			26: \hspace{12mm} Continue ($\tau_{PSM})$ \\
			27: \hspace{7mm} $i=i+1$  \\
	\end{tabu}}
\end{algorithm} 

\section{System configurations and Mathematical model of costs}\label{sec4}

In this section, all the input data or parameters, necessary for HOMER and Matlab simulation, are briefly described.

\subsection{System configurations}

\subsubsection{Solar PV}

\begin{figure}[H]
	\centering
	\includegraphics[width=\columnwidth, height=5cm]{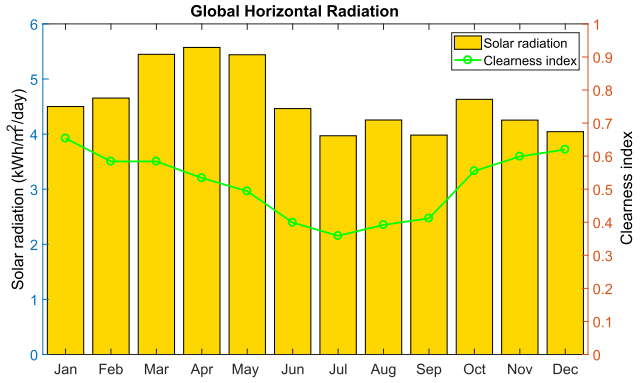}\\
	\caption{Solar radiation profile and clearness index of the study region.}			\label{profile}
\end{figure}

The average daily solar radiation profile of the selected area throughout a year has been illustrated in Figure~\ref{profile}. 

The total amount of energy generation using solar PV panels can be calculated by HOMER using the following equation  ~\cite{9022971, en11061500}:

\begin{equation}   
E_{PV} = R_{PV} \times PSH \times \eta_{PV}\times 365 \quad days/year,   \label{Eq:EPV}
\end{equation}

\subsubsection{BG}

The average biomass available in the selected area is 9 tons/day \cite{ISLAM2018338}. The output power of the BG can be expressed as follows \cite{CHAUHAN201499, ISLAM2018338}.

\begin{equation}  
P_{BG} = \frac{T_{BM} (t/year)\times CV_{BM}\times \eta_{BM} \times 1000}{365\times 860 \times t_{op}}.   \label{Eq:PBG}
\end{equation}

The amount of renewable energy harvested by the biomass generator can be expressed as follows \cite{CHAUHAN201499, ISLAM2018338}:

\begin{equation}   
E_{BG} = P_{BG}(365\times24\times C_f)  , \label{Eq:EBG}
\end{equation}

where, $C_f$ is the capacity factor that is the ratio between actual and maximum possible electrical energy. 

\subsubsection{Battery}

In this work, the Trojon L16P battery model has been used due to its large size, low cost, high reliability, low self-discharge, and low maintenance requirements. Battery lifetime ($L_{batt}$) is a directly related variable for the cost analysis, which can be calculated using the following equation \cite{su12229340}: 

{\begin{equation}
L_{batt} = min(\frac{N_{batt} \times T_{batt}}{T_{a}}, R_{batt})	,	\label{Eq:BLIFE}
\end{equation}

\subsubsection{Load Profile}

The proposed hybrid power supply system has focused to meet the power consumption of the BSs of a HetNet. On the basis of different applications, a HetNet, consisting different types of BSs, can optimise the Quality of Service (QoS) requirements. As a result, we consider a heterogeneous network. As a result, estimating the proper dimensions of macro, micro, pico, and femto-cell power consumption is important. The total power consumption of a BS is mainly dependent on traffic demand, we formulated the discrete power consumption considering 10MHz bandwidth from paper \cite{6056691}  using the following equation. Then on that basis, we design and optimise the overall power generation and sharing architecture considering energy, cost and environmental aspects. The assumed traffic demand profile for the proposed model is shown in Figure \ref{traffic}. 

\begin{equation}	\label{Eq:PIN}
P_{BS}=
\begin{cases}
N_{TRX}[P_{1}+\Delta_{p}P_{TX}(\chi - 1)], & \text{if}\ 0<{\chi}\leq 1 \\
N_{TRX}P_{sleep}, & \text{if}\ \chi=0,
\end{cases}
\end{equation}

where, ${P_{1}=P_{0} + \Delta_{p} P_{TX}}$.

\begin{figure}[H]
	\centering
	\includegraphics[width=.98\columnwidth, keepaspectratio]{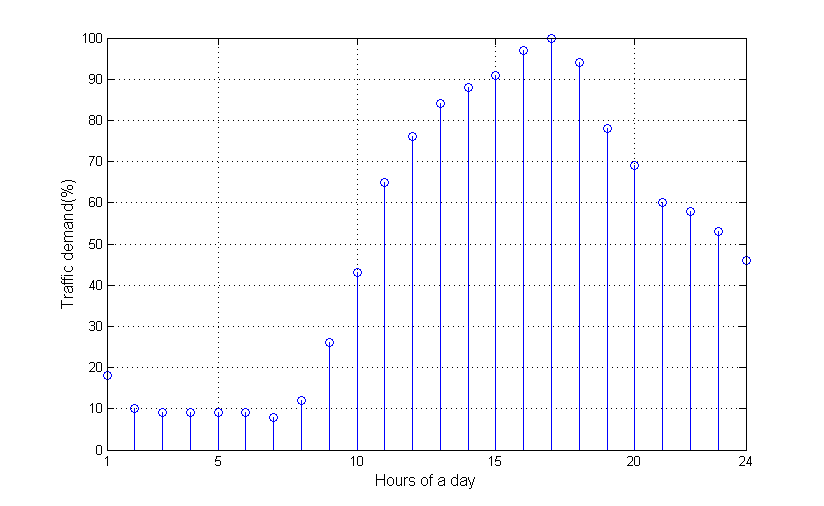}\\
	\caption{Assumed traffic profile for the simulation.}			\label{traffic}
\end{figure}

The discrete power consumption by the macro, micro, pico, and femto cellular LTE BSs considering 10 MHz bandwidth are obtained from ~\cite{6056691}. It implies that a smaller BS needs lower power due to the smaller radiated power and coverage area.

The power consumption of the BSs that is directly connected to the related loss factors, can be calculated as \cite{6056691, 6600717}:

\begin{equation} \label{Eq:P1}
{P_{1}} = \frac {{P_{BB}} + {P_{RF}} + {P_{PA}}}{({1-\sigma_{DC}}){(1-\sigma_{MS}}){(1-\sigma_{cool}})}
\end{equation}

Where, $P_{BB}$, $P_{PA}$ and $P_{RF}$ respectively refer to the baseband, power amplifier and RF power consumption scaled with the transmission bandwidth.

\subsection{BS Power Consumption Model}
\begin{table*}  []
	\centering
	\caption{BS power consumption at maximum load under 10 MHz bandwidth \cite{6056691}.}
	\label{Table:BSPOWER}
	{\tabulinesep=.5mm
		\begin{tabu}{|c|l|c|c|c|c|}
			\hline
			Components & Parameters  & Macro  & Micro & Pico & Femto \\ \hline
			& $P_{max}$ [W] & 20 & 6.3 & 0.13 & 0.05 \\
			& Feeder loss $\sigma_{feed}$ [dB] & 0 & 0 & 0 & 0 \\ \hline
			\textbf{PA} & Back-off [dB] & 8 & 8 & 12 & 12 \\
			& Max PA out [dBm] & 51 & 46 & 33 & 29 \\
			& PA efficiency $\eta_{PA}$ [\%] & 31.1 & 22.8 & 6.7 & 4.4\\ \hline
			& \textbf{Total PA}, $P_{PA}$ [W]& 64.4 & 27.7 & 1.9 & 1.1 \\\hline
			\textbf{RF} & $P_{TX}$ [W] & 6.8 & 3.4 & 0.4 & 0.2\\
			& $P_{RX}$ [W] & 6.1 & 3.1 & 0.4 & 0.3 \\ \hline
			& \textbf{Total RF}, $P_{RF}$ [W] & 12.9 & 6.5 & 1.0 & 0.6\\ \hline
			\textbf{BB} & Radio (inner Rx/Tx) [W] & 10.8 & 9.1 & 1.2 & 1.0 \\
			& Turbo code (outer Rx/Tx) & 8.8 & 8.1 & 1.4 & 1.2 \\
			& Processors [W] & 10 & 10 & 0.4 & 0.3\\ \hline
			&\textbf{Total BB}, $P_{BB}$ [W] & 29.6  & 27.3 & 3.0 & 2.5 \\\hline
			\textbf{DC-DC} & $\sigma_{DC}$  [\%] & 7.5  & 7.5 & 9 & 9\\
			\textbf{Cooling} & $\sigma_{cool}$  [\%] & 0 & 0 & 0 & 0 \\
			\textbf{Main Supply} & $\sigma_{MS}$  [\%] & 9 & 9 & 11 & 11\\ \hline
			Sectors  & & 3 & 1 & 1 &1 \\
			Antennas &  & 2 & 2 & 2 &2\\ \hline
		\textbf{Total [W]} &  & \textbf{754.8} & \textbf{144.6} & \textbf{14.7} & \textbf{10.4}\\ \hline  
	\end{tabu}}
\end{table*}

\begin{table}
	\centering
	\caption{BS key parameters \cite{6056691}.}
	\label{Table:BSSLEEP}
	{\tabulinesep=0.3mm
		\begin{tabu}{|l|c|c|c|c|c|} \hline
  BS Type     & $N_{TRX}$ & $P_{MAX}$ [W] & $P_0$ [W] & $\Delta_{p}$ & $P_{Sleep}$ [W] \\ \hline
  Macro & 6         & 20            & 84        & 2.8          & 56 \\ \hline
  Micro  & 2         & 6.3           & 56        & 2.6          & 39 \\ \hline
  Pico    & 2         & 0.13           & 6.8      & 4.0         & 4.3\\ \hline
  Femto    & 2         & 0.05           & 4.8      & 8.0        & 2.9\\ \hline
		\end{tabu}}
\end{table}

%

\subsubsection{Simulation Setup for HOMER}
We first study the techno-economic feasibility of the proposed system. In the simulation setup, we assume a project duration is 20 years and the yearly interest rate is 6.75\% \cite{Bank}. HOMER platform has been used for this study. The system architecture of the proposed heterogeneous network is shown in Figure \ref{Architecture}. The initial values of the main parameters have been listed in Table \ref{Table:7}.

\begin{table}[H]
	\centering
	\caption{Key parameters and their values for HOMER simulation setup~\cite{ISLAM2018338} \cite{CHAUHAN201499}.}	\label{Table:7}
	{\tabulinesep=0.5mm
		\begin{tabular}{clc}
			\toprule
			\textbf{Components} & \textbf{Parameters} & \textbf{Value} \\ \midrule
			Resources & Solar intensity & 4.59 kWh/m$^2$/day \\
			& Biomass available & 9 t/day  \\
			& Interest rate & 6.75\%     \\ \midrule	
			Solar PV & Operational lifetime & 25 years \\
			& Derating factor & 0.9 \\
			& System tracking & Dual-axis \\
			& CC & \$1/W  \\ 
			& RC & \$1/W     \\
			& OMC/year & \$0.01/W		\\ \midrule
			BG   & Efficiency  & 30\% \\
			& Operational lifetime & 25,000 h\\
			& CC & \$0.66/W \\
			& RC & \$0.66/W \\
			& OMC/year &  \$0.05/h \\
			& FC &  \$30/t     \\ \midrule  
			Battery & Round trip efficiency & 85\% \\ 
			& $B_{SoC_{min}}$ & 30\% \\
			& $V_{nom}$   & 6 V   \\
			& $Q_{nom}$		& 360 Ah	\\
			& CC & \$300/unit   \\
			& RC & \$300/unit  \\
			& OMC/year &  \$10/unit  \\ \midrule
			Converter & Efficiency & 95\%\\
			& Operational lifetime & 15 years \\
			& CC  & \$0.4/W    \\
			& RC & \$0.4/W   \\
			& OMC/year &  \$0.01/W    \\  \bottomrule 	
			\end{tabular}}
\end{table}

\begin{figure}[htb]
	\begin{multicols}{2}
		\includegraphics[width=\linewidth, height= 4.5cm]{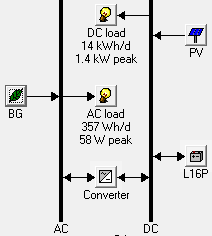}\par \includegraphics[width=\linewidth, height= 4.5cm]{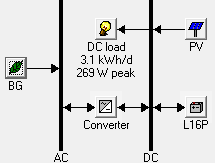}\par
	\end{multicols}
	\caption{System architecture in the HOMER platform for macro (left) and micro (right) BSs.}
	\label{Architecture}
\end{figure}

\begin{figure}[htb]
	\begin{multicols}{2}
		\includegraphics[width=\linewidth]{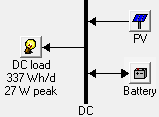}\par
		\includegraphics[width=\linewidth]{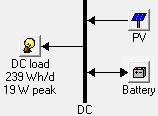}\par
	\end{multicols}
	\caption{System architecture in the HOMER platform for pico (left) and femto (right) BSs.}
	\label{Architecture}
\end{figure}

\subsubsection{Simulation Setup for Resource scheduling}

Monte-Carlo simulation parameters has shown in Table \ref{RB}.

\begin{table*}[htb]
	\centering
	\caption{Key parameters for MATLAB-based Monte-Carlo simulation setup.\cite{3GPP2011}}
	\label{RB}
	{
		\begin{tabular}{l l}
			\toprule
			\textbf{Parameters} & \textbf{Value} \\
			\midrule
			Resource block (RB) bandwidth & 180kHz \\
		    System bandwidth, BW & 5, 10, 15, 20 MHz\\
			Carrier frequency, $f_c$ & 2 GHz\\
			Duplex mode & FFD\\
			Cell radius & 1000m (macro), 500m (micro), 200m (pico), 10m (femto)\\
			BS Transmission power & 20W (macro), 6.3W (micro), 0.13W (pico), .05W (femto) \\
			Noise power density & -174dBm/Hz\\
			Path loss exponent, $\alpha$ & 3.574 \\
			Shadow fading $X_{\sigma}$ & 8 dB\\
			Access technique, DL & OFDMA \\
			Traffic distribution & Randomly distributed \\
			\bottomrule	
	\end{tabular}}
\end{table*}

For the simulation setup of Algorithm \ref{Algorithm1}, the assumed value of the parameters are shown in Table \ref{NRT}. 

\begin{table*}[htb]
	\centering
	\caption{Simulation parameters for Algorithm \ref{Algorithm1}}
	\label{NRT}
	{
		\begin{tabular}{l l}
			\toprule
			\textbf{Parameters} & \textbf{Value} \\
			\midrule
			$\chi$(Packet arrival rate)(M/G/1 stochastic process) & 0.05 to 0.5 packets/ms \\
		    $\mu$ (Service rate) & 100 packets/ms\\
			$\tau_S$(Short cycle timer length) & 20ms\\
			$\tau_L$(Long cycle timer length) & 320ms\\
			$\tau_{st}$(Step size timer) & 20ms\\
			$\tau_i$ (Inactivity timer) & 10ms\\
			$N$(No of short cycles) & 1 to 16\\
			$P_{Data}$(Power consumption during ON period) & 500mW \\
			$P_{Inactive}$(Power consumption during inactivity timer) & 255mW\\
			$P_{OFF}$ (Power consumption during DRX period) & 11mW \\
			\bottomrule	
	\end{tabular}}
\end{table*}

The parameters for the simulation setup of Algorithm \ref{Algorithm2} are listed in Table \ref{Table:1}.  

\begin{table*}[htb]
	\centering
	\caption{Simulation parameters for Algorithm \ref{Algorithm2}}
	\label{Table:1}
	{
		\begin{tabular}{l l }
			\toprule
			\textbf{Parameters} & \textbf{Value} \\
			\midrule
			$\chi$(Packet arrival rate)(M/G/1 stochastic process) & 0.005 to 0.5 packets/ms \\
			$\mu$ (Service rate) & 100 packets/ms\\
			$\tau_{eDRX}$(known as $T_{3324}$)(eDRX long cycle timer length) & 5:2:35\\
			$\tau_{PSM}$(known as $T_{3412}$)(PSM sleep timer) & 3 hours\\
			$\tau_{i}$ (Inactivity timer) & 10ms\\
			$N$(No of short cycles) & 1 to 16\\
			$P_{Data}$(Power consumption during ON period) & 500mW \\
			$P_{Inactive}$(Power consumption during inactivity timer) & 255mW\\
			$P_{eDRX}$ (Power consumption during eDRX period) & 11mW \\
			$P_{PSM}$(Power consumption during deep sleep period) & 0.0108$\mu$W\\
			\bottomrule	
	\end{tabular}}
\end{table*}

\section{Results Analysis}\label{sec5}

The simulation output from HOMER was thoroughly analyzed using MATLAB simulation from the following key perspectives: i) Optimization standard, (ii) Energy Generation and Consumption, (iii) Energy sharing, (iv) Economic issue, (v) Energy efficiency issue, (vi) Power saving issue for NRT and IoT devices, and (vii) Carbon footprint issue. 

\subsection{Optimization Standard}
An optimal size of the components used for the proposed hybrid RES for the HetNet is summarized in Table \ref{Table:8}. The proposed system is based on off-grid energy sharing and in terms of the consumed energy, the macro, and micro BSs will be served by solar PV and BG with battery storage and converter. Due to small energy requirements, pico and femto BSs will only be aided by solar PV and backup storage devices and as a consequence, no converter is needed to support the DC load. The proposed model has been simulated on the HOMER platform taking the dynamic behavior of traffic and RES into account. It is recognized that a higher value of the component size is required for the bigger cell sizes due to the higher amount of energy consumption. With the change of system bandwidth, the optimal size of the BG, battery units, and converters are remaining unchanged. On the other hand, the optimal size of the solar PV increase with the increment of system BW for both macro and micro BS configurations to cope up with the higher energy demand. As such,  the proposed system is  compatible under dissimilar BS network requirements.

\begin{table*}[hbt]
	\centering
	\caption{An optimal size of the hybrid solar PV/BG system.}
	\label{Table:8}
	\resizebox{\textwidth}{!}
		{\tabulinesep=0.4mm
			\begin{tabular}{ccccccccccccccccc} 
				\toprule
				\textbf{BW}  &  \multicolumn{4}{c}{\textbf{PV (kW)}}&  \multicolumn{4}{c}{\textbf{BG (kW)}} &\multicolumn{4}{c}{\textbf{Battery(units)}}   &  \multicolumn{4}{c}{\textbf{Converter (kW)}}  \\  \cmidrule{1-17}
				\textbf{(MHz)}  &\textbf{Macro}  &  \textbf{Micro} &\textbf{Pico}  &  \textbf{Femto}  & \textbf{Macro}  &  \textbf{Micro} &\textbf{Pico}  &  \textbf{Femto}  &\textbf{Macro}  &  \textbf{Micro} &\textbf{Pico}  &  \textbf{Femto}  &\textbf{Macro}  &  \textbf{Micro} &\textbf{Pico}  &  \textbf{Femto}  \\\midrule
				5  & 2.5 & 0.5 & 0.1 & 0.1 & 1 & 1 & * & * & 32 & 24 & 8 & 8 & 0.1 & 0.1 & * & * \\ 
				10  & 3.5 & 0.5 & 0.1 & 0.1 & 1 & 1 & * & * & 32 & 24 & 8 & 8 & 0.1 & 0.1 & * & * \\
				15  & 4 & 1 & 0.1 & 0.1 & 1 & 1 & * & * & 32 & 24 & 8 & 8 & 0.1 & 0.1 & * & * \\
				20  & 4 & 1 & 0.1 & 0.1 & 1 & 1 & * & * & 32 & 24 & 8 & 8 & 0.1 & 0.1 & * & * \\ \bottomrule 
	\end{tabular}}
\end{table*}

\subsection{Energy Generation and Consumption}

The yearly energy consumption (DC and AC load) including excess energy for the macro, micro, pico, and femto-cell networks under 10MHz bandwidth are summarised in Table \ref{Table:A}. This excess energy is used as a backup, either for charging the battery or for sharing in need of other BSs. DC load is the main energy consumer. It is assumed that a bulb as an AC load is only connected macro BS.

\begin{table*}[hbt]
	\centering
	\caption{Energy consumption breakdown for 10 MHz system bandwidth.}
	\label{Table:A}
	{
		\begin{tabular}{l c c c c}
			\toprule
			\textbf{Items} & \textbf{Macro} & \textbf{Micro} & \textbf{Pico} & \textbf{Femto}\\
			\midrule
			AC load (kWh/yr) & 130 (3\%) & 0 (0\%) & 0 (0\%) & 0 (0\%)\\
			DC load (kWh/yr) & 4,989 (97\%) & 1,117 (100\%) & 123 (100\%) & 87.2 (100\%) \\
			\midrule
			Excess energy (kWh/yr) & 2,056 (26.8\%) & 128 (9.41\%) & 81.3 (37.6\%) & 121 (55.9\%) \\
			\bottomrule	
	\end{tabular}}
\end{table*}

The individual energy generation breakdown of the proposed off-grid hybrid system is illustrated in Figure \ref{Pi}. For the simulation, the dynamic profile of the RES (Figure \ref{profile} and Figure \ref{solar}) has been considered. Here results are only shown for 10MHz system BW. For all cases, the solar PV is the main energy contributor. In the proposed model, macro BSs can share energy to another macro, micro, pico, and femto BSs. Consequently, more excess electricity has been generated in the case of macro BSs. The rest of the BSs use the excess electricity as a backup in case of a shortage or outage of RES. In case of shortage, these BSs can share excess electricity only to the mother macro BS.

\begin{figure}[H]
	\centering
	\includegraphics[width=\columnwidth]{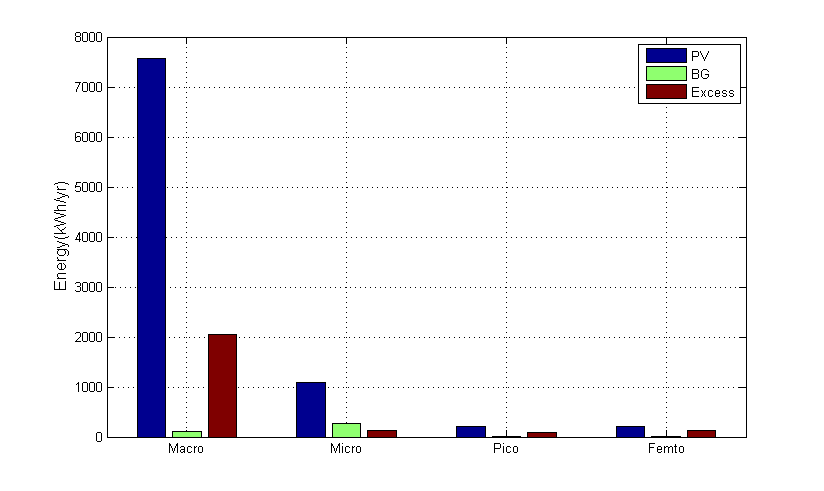}\\
	\caption{Individual energy breakdown for 10 MHz system bandwidth.}			
	\label{Pi}
\end{figure}

\begin{figure*} [htb]
	\begin{multicols}{2}
		\includegraphics[width=\linewidth]{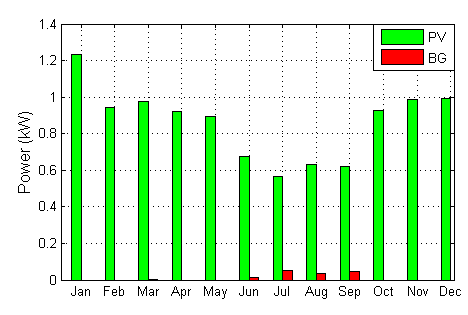}\par 
		\includegraphics[width=\linewidth]{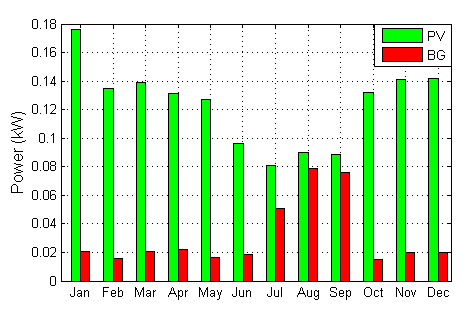}\par 
	\end{multicols}
	\caption{Monthly power contribution by the RES under 10 MHz for macro (left) and micro (right) BS.}
	\label{A}
\end{figure*}

\begin{figure*}[htb]
	\begin{multicols}{2}
		\includegraphics[width=\linewidth]{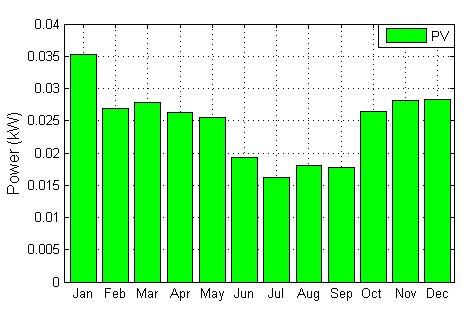}\par 
		\includegraphics[width=\linewidth]{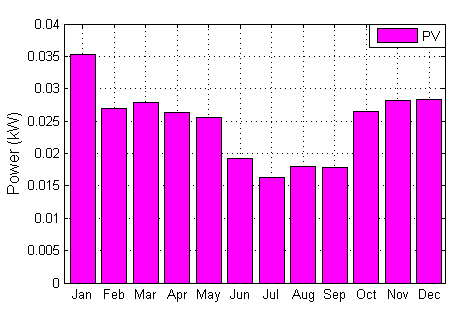}\par 
	\end{multicols}
	\caption{Monthly power contribution by the RESs under 10 MHz for pico (left) and femto (right) BS.}
	\label{B}
\end{figure*}

The periodic monthly statistics of the power generation by the different RES for different BSs are separately illustrated in Figure \ref{A} and Figure \ref{B}. Solar PV is the main contributor throughout the year and BG contributes to that period when solar intensity is relatively low. It is clear that the proposed system can minimize the capital costs and carbon contents because the solar PV generates a large of consumed energy for the overall heterogeneous setup. On the other hand, the system is also reliable because of the presence of BG and the energy sharing technique. 

\begin{figure*}[htb]
	\begin{multicols}{2}
		\includegraphics[width=\linewidth]{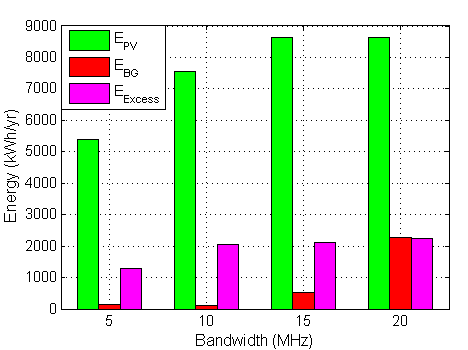}\par 
		\includegraphics[width=\linewidth]{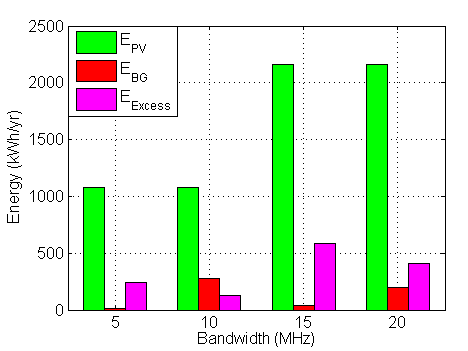}\par 
	\end{multicols}
	\caption{Energy breakdown under different system BW for the macro (left) and micro (right) BS.}
	\label{C}
\end{figure*}

\begin{figure*}[htb]
	\begin{multicols}{2}
		\includegraphics[width= \linewidth]{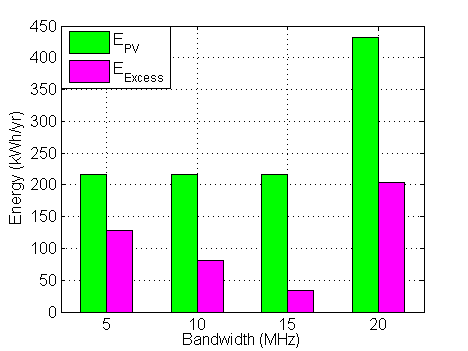}\par 
		\includegraphics[width= \linewidth, height=6.3cm]{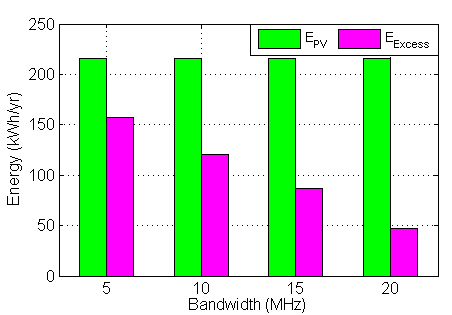}\par 
	\end{multicols}
	\caption{Energy breakdown under different system BW for the pico (left) and femto (right) BS.}
	\label{D}
\end{figure*}

A broad evaluation of the annual energy contributed by the different sources under various system BWs has been illustrated in Figure \ref{C} and and Figure \ref{D}, where it is clear that with an increase in  system bandwidth, the generation is increased. If solar PV cannot bear the load because of cost efficiency, BG will take the place. In the case of pico and femto-cells, with the increment of the system BW, the excess electricity will  decrease to cope up with the load. It is better for the system performance, as the setup will not change under different system BWs. 

The thorough calculation of yearly energy generation under 10 MHz BW is given below:

\textbf{Solar PV}

To generate a target maximum power ($P_P$), the required number of PV panel can be expressed as:

$N_{PV}$=$\frac{P_p}{P_{nom}}$=$\frac{3.5kW}{250W}$=14

where $P_{nom}$ is the nominal power of the solar PV panel. 

From the above calculation, the required number of solar PV module is 14: as connected 2 in series and 7 in parallel.

The annual energy generated by the 3.5 kW solar PV module can be determined using equation \ref{Eq:EPV} in the following way:

$E_{PV}$=3.5 kW $\times 4.59 kWh/m^2/day \times  0.9 \times 365$ days/year = 5277.35 kWh/yr.

Additionally, a dual-axis tracking mode of solar PV module increases the total amount of energy by 43.4\% resulting in 7,567 kWh/yr.\\

\textbf{Biomass generator}

The annual energy generated by the 1 kW BG can be determined using (\ref{Eq:PBG}) and (\ref{Eq:EBG}) as follows:

$P_{BG} = \frac{0.149 \text{ton/year} (T_{BM})\times 3411.33 \text{kCal/kg} (CV_{BM})\times 0.30(\eta_{BM}) \times 1000}{365\times 860 \times 0.9726 (t_{op})}$
= $0.4995$kW

$E_{BG}=P_{BG}(365\times24\times C_f)=0.4995\times 365\times24\times 0.0245= 107\text{kWh/yr}$\\

\textbf{Battery bank}




The nominal rated voltage of the Trojan L16P battery is 6V; hence, a 48V DC bus–bar connected in series is used. As a consequence, the total of units is always multiple of 8 batteries.

\begin{figure}[htb]
	\centering
	\includegraphics[width=\columnwidth, height=6.1cm]{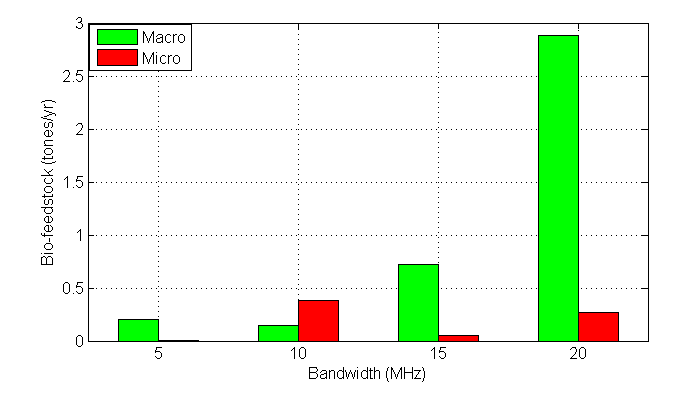}\\
	\caption{Bio-feedstock consumption.}				\label{Feedstock}
\end{figure}

\begin{figure}[htb]
	\centering
	\includegraphics[width=\columnwidth]{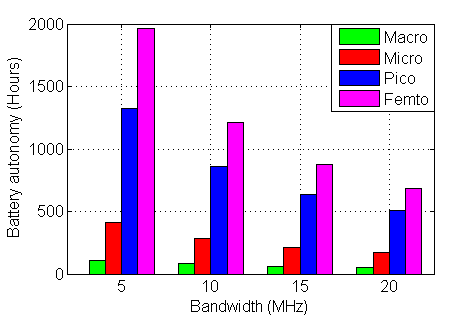}\\
	\caption{Battery autonomy under different system bandwidth.}			\label{Autonomy}
\end{figure}

\begin{table*}[hbt]
	\centering
	\caption{Annual energy savings due to energy sharing mechanism between two adjacent macro BSs.}
	\label{Table:9}
	{\tabulinesep=0.5mm
		\begin{tabular}{cccccccc}
			\toprule
			{\textbf{BW (MHz)}} & {\boldmath{$E_{gen}$ \textbf{(kWh)} }}& {\boldmath{$E_{D}$ \textbf{(kWh)} }}&  {\boldmath{$E_{excess}$ \textbf{(kWh)} }}&  {\textbf{I (Amp)}} &{\boldmath{$E_{loss}$ \textbf{(kWh)}}}   &  {\boldmath{$E_{share}$\textbf{ (kWh)}}} &  {\boldmath{$E_{save}$ \textbf{(\%)}}} \\  \midrule
			5  & 5548 & 3896 & 1295  & 3.07 & 470.87 & 824.12 & 21.15\\ 
			10  & 7674 & 5120 & 2056 & 4.88 & 1187.20 & 868.79 & 16.96\\
			15  & 9167 & 6434 & 2116 & 5.03 & 1256.67 & 859.32 & 13.35\\
			20  & 10808 & 7941 & 2237  & 5.32 & 1405.75 & 831.24 & 10.46\\ \bottomrule 
	\end{tabular}}
\end{table*}

\begin{table*}[hbt]
	\centering
	\caption{Amount of shared energy and percentage of energy-saving for inter BS sharing mechanism under 10 MHz.}
	\label{Table:10}
	{\tabulinesep=.6mm
		\begin{tabular}{ccccc}
			\toprule
			\textbf{Sharing}    & \boldmath{$E_{excess}$ \textbf{(kWh)}} & \textbf{Resistance ($\Omega$)}   & \boldmath{$E_{loss}$ \textbf{(kWh)}}   &\boldmath{$E_{share}$\textbf{ (kWh)}}     \\ \midrule
			\textbf {Macro-Macro}   & 2056  & 5.67    & 1187.20        & 868.79     \\
			\textbf {Micro-Macro}  & 128  & 1.70    & 1.34            & 126.66    \\
            \bottomrule      
	\end{tabular}}
\end{table*}

\subsection{Economic issue}

Bio-feedstock is the raw material of BG. In our proposed system, only macro and micro BSs are connected to both Solar PV and BG, so liable for bio-feedstock consumption. Bio-feedstock consumption is linearly proportional to the amount of energy harvested from the BG and for our proposed model the annual consumption has shown in Figure \ref{Feedstock}. It is clear from the previous energy breakdown figures that total energy consumption has proportionally been increased with the increases  of BW but BG is not the only source for the energy generation of macro and micro BSs. As a result, bio-feedstock consumption has not proportionally increased with the BW though macro BSs average bio-feedstock consumption is far more than micro BSs.

The requirement of total battery for the proposed model has increased with the growth of total load. As macro BSs has to deal with the highest load, so a total 32 of batteries for the energy storage purpose. The number of batteries required for micro, pico, and femtocells are 16, 8, and 8 respectively to backing the load request of the BS throughout the shortage or absence of RES. The number of batteries are not feasible less than 8 because the proposed model use the “Trojan L16P” battery. The DC bus bar voltage is 48V. The terminal voltage of each battery is 6V. HOMER calculates the battery bank autonomy using equation (\ref{Eq:BAUT}). The annual battery bank autonomy under different network configurations are illustrated in Figure \ref{Autonomy}. The battery bank autonomy is inversely proportional to the BS load or energy demand. For the same reason, higher system bandwidth reveals lower battery autonomy. For a reliable and cost-effective supply system, these results are justified.

\begin{figure*}[h]
	\begin{multicols}{2}
		\includegraphics[width=\linewidth, height=5cm]{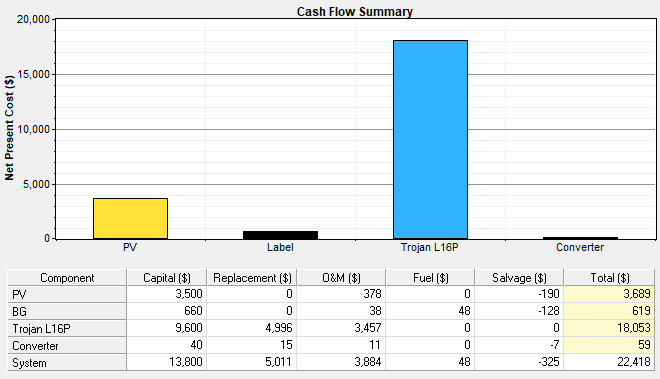}\par 
		\includegraphics[width=\linewidth, height=5cm]{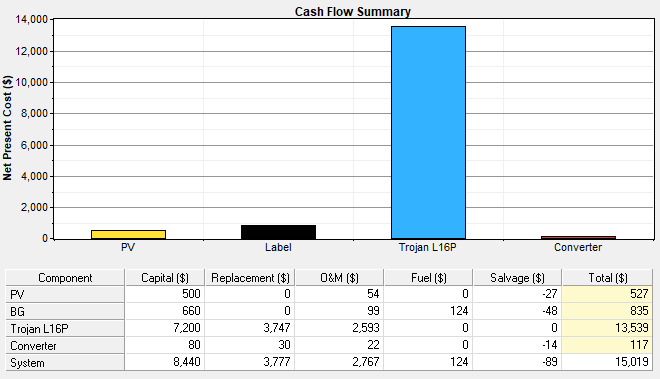}\par 
	\end{multicols}
	\caption{Cash flow summary of the proposed system for macro (left) and micro (right) BS.}
	\label{E}
\end{figure*}

\begin{figure*}[htb]
	\begin{multicols}{2}
		\includegraphics[width=\linewidth, height=5cm]{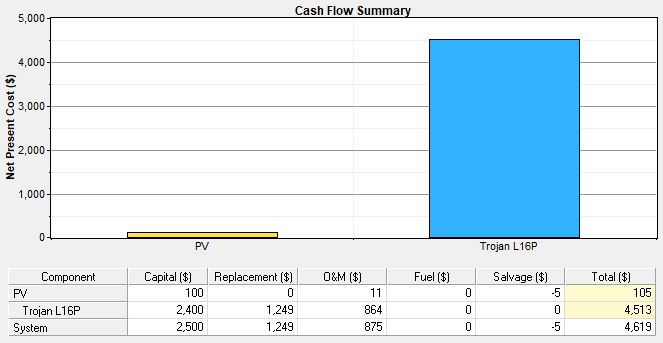}\par 
		\includegraphics[width=\linewidth, height=5cm]{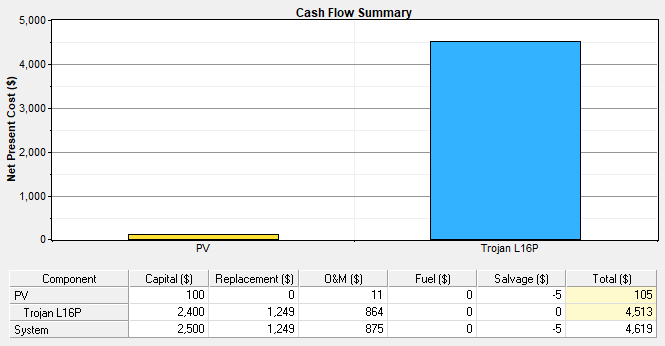}\par 
	\end{multicols}
	\caption{Cash flow summary of the proposed system for pico (left) and femto (right) BS.}
	\label{F}
\end{figure*}

\subsection{Energy Sharing}

According to the flow diagram in Figure \ref{Energy_sharing}, excess energy has been shared through a low resistive physical line among the adjacent BSs. The value of the resistance of the interconnected line has been taken from the American Wire Gauge (AWG) standard conductor size table which is 3.276 $\Omega$/km \cite{Solaris}. The inter-cell distance is calculated as $\sqrt 3$ times of cell radius (i.e., $\sqrt 3$R) and the cell radius is 1000m for $P_{TX}$ = 20 W. The total resistance of the transmission line between the two BSs is 5.67 $\Omega$. We use the cell radius of macro, micro, pico and femto BSs as 1000m, 500m, 200m and 10m, respectively. This energy sharing among the BSs for the heterogeneous network ensures the optimum use of RES and avoids any type of blackout or power shortage. The percentage of energy savings with the total amount of shared energy for different system bandwidth and sharing policies is demonstrated in Table \ref{Table:9} and Table \ref{Table:10} respectively. The results show that around 10\% to 21\% of total energy can be saved every year by applying the proposed macro-macro energy-sharing technique. The proposed system ensures energy efficiency ($\eta_{EE}$) under dynamic traffic profile and can save a considerable amount of energy loss by energy sharing technique.

A critical analysis of associated costs to the proposed system has represented in this section. These analyses have been done via HOMER optimisation software to propose a reliable and optimum cost-efficient green supply system. Nominal cash flow summary for the proposed system of macro, micro, pico, and femto BSs have respectively shown under 10MHz system BW in Figures \ref{E} and \ref{F}. For all the demonstrated configurations, the capital cost (CC) has the highest value and replacement cost (RC) has the second-highest value. For macro BS, CC for the PV array is \$3,500 (panel size 3.5 kW, cost \$500/0.5 kW), operation and maintenance cost (OMC) is \$378 (panel size 3.5 kW, cost around \$10/1 kW), RC is zero because the PV array has a lifespan of 25 years which is prolonged than the project lifespan. The salvage value for that extra five years is \$190. For other BSs, different costs of PV array have been calculated in the same fashion. For both macro and micro BS configurations, a single BG is required, which costs \$660. OMC and fuel cost (FC) vary on generation. The salvage value is inversely proportional to the BG use. The number of batteries is always multiple of 8 because the DC bus bar voltage is 48V (Each battery's nominal voltage is 6V). CC for each battery is \$300. RC depends on battery lifetime. Battery lifetime can be calculated from equation \ref{Eq:BLIFE}. RC of each battery is also \$300. Less use of battery extends the battery lifetime. Higher battery lifetime minimizes the overall RC of the battery for the system. OMC of the battery is \$10/unit/year. So 32 batteries cost \$320 per year. But due to the discount factor, this costs less in the long run. CC, RC, and OMC of converter depend upon the quantity of use. CC and RC are both \$.4/Watt and OMC is \$.01/Watt/year. More BG generation costs more for the converter. For pico and femto BSs, there is no BG need for the system. As a consequence, no converter is needed either. In comparison with other components, the battery bank is responsible for the highest amount of CC, RC, and OMC in all configurations. 

\begin{figure}[H]
	\centering
	\includegraphics[width=\columnwidth]{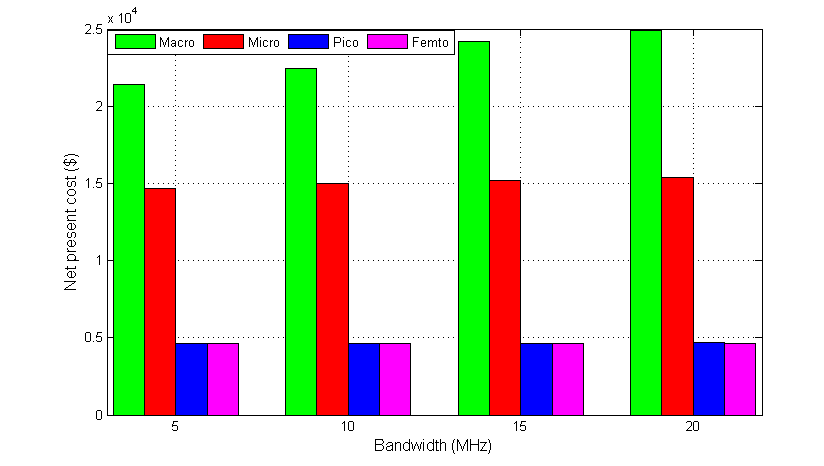}\\
	\caption{NPC under different system bandwidth.}			
	\label{G}
\end{figure}

\begin{figure}[H]
	\centering
	\includegraphics[width=\columnwidth]{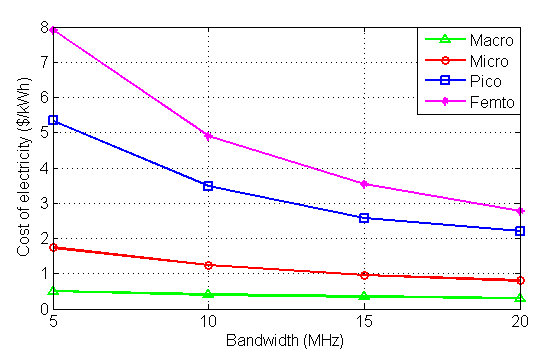}\\
	\caption{CoE under different system bandwidth.}			
	\label{H}
\end{figure}

Cost is linearly proportional to the load or energy demand. A widespread comparison of the NPC and CoE under different network configurations has been illustrated in Figure \ref{G} and Figure \ref{H} respectively. With the increment of system BW, load or energy demand has increased. As a consequence, a clear surge has been seen in NPC. In contrast, a higher rate of service ensures lower CoE. In terms of NPC, pico and femtocell configurations have shown identical value due to low load but identical system requirements. But, the excess electricity has been used automatically for sharing (only for macro and micro) or storage purposes based on the real-time requirement.  As a result, the CoE is not same in case of pico and femto BSs.


\subsection{Energy Efficiency issue}

User throughput and the system’s energy efficiency are two vital variables in the case of performance analysis of the system model. Daily throughput variation under different BS configurations has been illustrated in Figure \ref{Throughput} which is in line with the traffic intensity. To evaluate throughput and energy efficiency performance under different system bandwidth a two-tier LTE cellular network has been considered. With the increment of system bandwidth, both throughput and energy efficiency has increased. With the increment of cell size, the required throughput has been achieved for all the cases. In the case of energy efficiency, the small cell has shown a better output. These performances are illustrated in Figure \ref{Through} and Figure \ref{EE}. 

\begin{figure}[htb]
	\centering
	\includegraphics[width=\columnwidth, height=7cm]{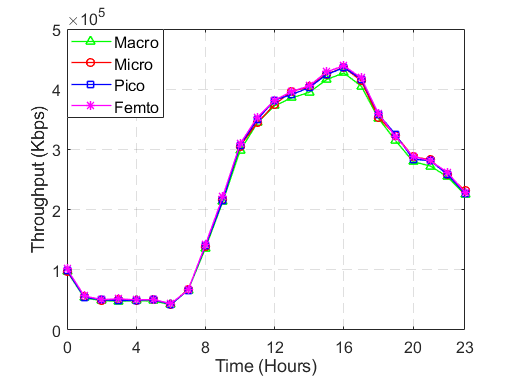}\\
	\caption{Throughput performance for two-tier HETNET.}
	\label{Throughput}
\end{figure}

\begin{figure}[htb]
	\centering
	\includegraphics[width=\columnwidth]{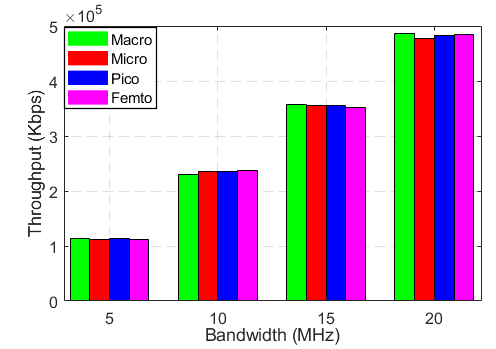}\\
	\caption{Throughput performance under different system bandwidths.}			\label{Through}
\end{figure}

\begin{figure}[htb]
	\centering
	\includegraphics[width=\columnwidth]{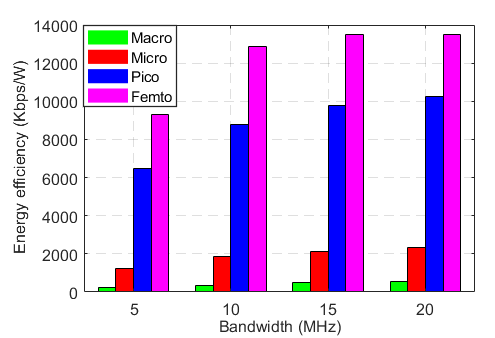}\\
	\caption{Energy efficiency under different system bandwidths.}			\label{EE}
\end{figure}

\subsection{Power saving issue for NRT and IoT devices}
\begin{figure}[htb]
	\centering
	\includegraphics[width=\columnwidth]{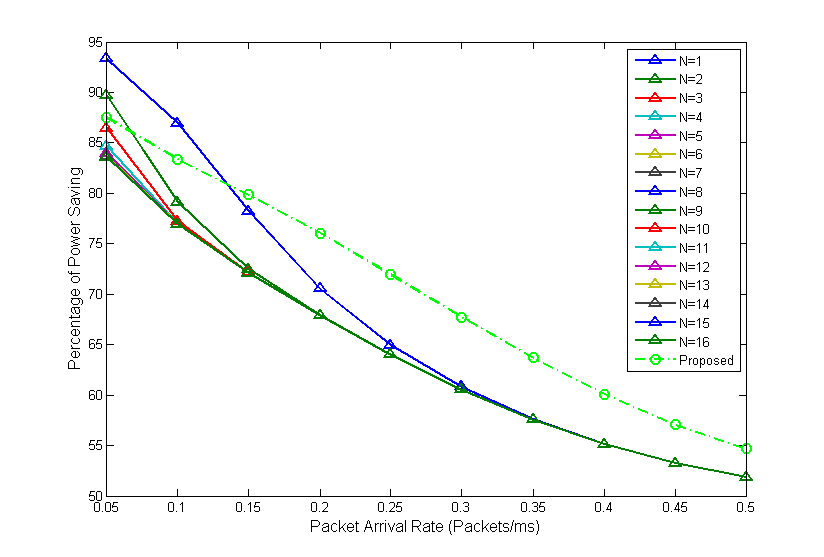}\\
	\caption{Percentage of power-saving over packet arrival rate (Packets/ms).}			\label{Power}
\end{figure}

\begin{figure}[htb]
	\centering
	\includegraphics[width=\columnwidth]{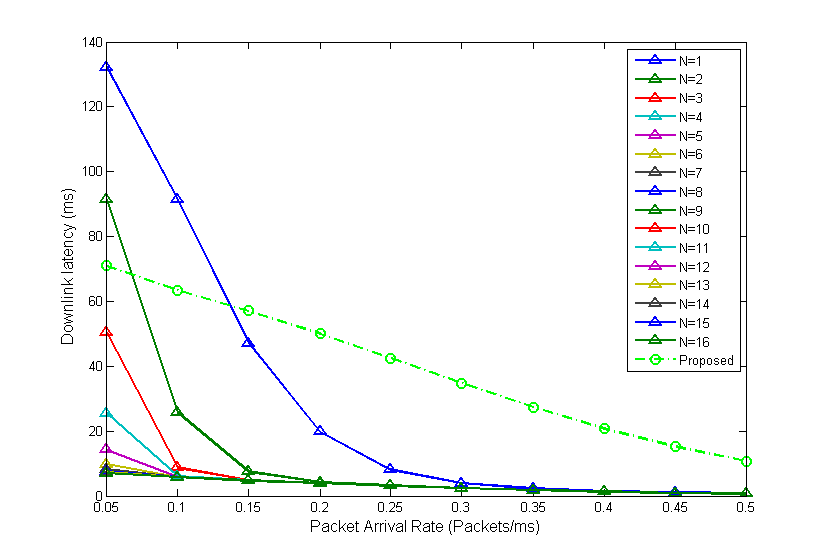}\\
	\caption{Downlink latency over packet arrival rate (Packets/ms).}			\label{Latency}
\end{figure}

Applying  the  proposed  resource  scheduling  algorithm  for NRT  applications,  the  system  has  achieved  better  power saving with a permissible delay which has shown in Figures \ref{Power} and \ref{Latency}. Although for N=1 , when no. of the short cycle is 1, the existing model shows better power-saving only for a very low packet arrival rate. But for that specific case, the downlink latency does not permit resource allocation. In all other cases, our proposed scheduling model shows a better performance than the existing model. Overall, it is seen that there is a trade-off between the downlink latency (delay) and the rate of power-saving.
\begin{figure}[htb]
	\centering
	\includegraphics[width=\columnwidth]{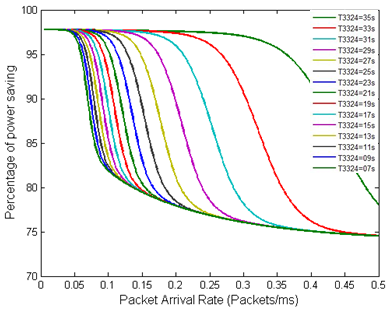}\\
	\caption{Percentage of Power Saving vs. packet arrival rate (IoT Applications).}			
	\label{Battery}
\end{figure}
On the other hand, simulating the proposed eDRX and PSM assisted resource scheduling algorithm for IoT applications, the percentage of power-saving over packet arrival rate is shown in Figure \ref{Battery}. From Figure \ref{Battery} it is clearly seen that, the more eDRX sleep mode(T3324) timer runs, the more power will save.    

\subsection{Carbon Footprints Issue}
\begin{figure}[htb]
	\centering
	\includegraphics[width=\columnwidth]{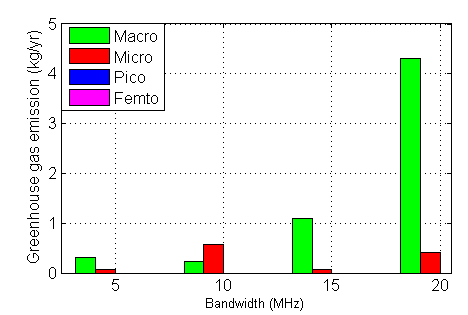}\\
	\caption{GHG emission for different system bandwidth.}		
	\label{GHG}
\end{figure}

To make a sustainable future, one of the sustainable development goals is known as SDG7 which targets affordable and clean energy. The proposed energy generation is dependent mostly upon solar PV, which does not discharge any carbon content. Although to achieve reliability, BG has added to the system in case of macro and micro BS, but it can minimize the carbon emission indirectly by reducing the burning of rice husk as shown in Table \ref{Table:13}. It is worth mentioning that rice husk can be treated as the primary resource of biomass which is generally used for cooking purposes in proposed rural areas. However, generation from BG is directly proportional to the $CO_2$ emission which has demonstrated in Figure \ref{GHG}. This figure shows the GHG emissions produced by the macro, micro, pico, and femto BSs under dissimilar system bandwidth. With the increment of system BW, the data of emission showing a clear proportional result with the bio feedstock consumption for BG. The proportional increment of GHG based on the operating hours of BG has shown in Figure \ref{BG}. The emission of total carbon contents from the proposed model under 10MHz BW are shown in Table \ref{Y}. 

\begin{figure}[htb]
	\centering
	\includegraphics[width=\columnwidth, height=5.6cm]{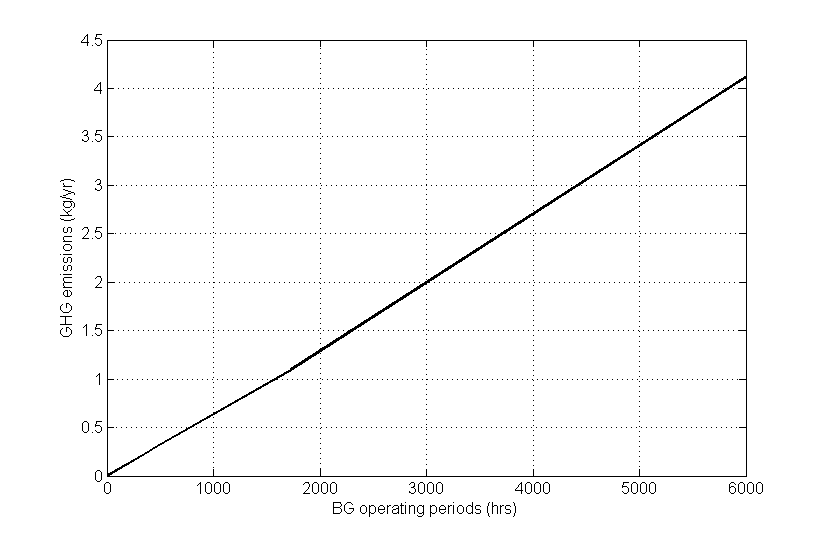}\\
	\caption{GHG emissions concerning BG operating hours.}			
	\label{BG}
\end{figure}

\begin{table}[H]
	\centering
	\caption{Carbon contents for the hybrid solar PV/BG system under 10 MHz bandwidth.}
	\label{Y}
	{\tabulinesep=0.5mm
		\begin{tabular}{lcc}
			\toprule
			\textbf{Pollutants} &  \multicolumn{2}{c}{\textbf{Emissions (Kg/yr)}}   \\  \cmidrule{2-3}
			&	\textbf{Macro}  &  \textbf{Micro} \\\midrule
			$CO_2$                & 0.216   & 0.558  \\ 
			$CO$                  & 0.000967   & 0.00249 \\
			Unburned hydrocarbons & 0.000107   & 0.000276 \\
			Particulate matter    & 0.0000729 & 0.000188 \\
			$SO_2$                & 0         & 0 \\
			$NO, NO_2, N_2O, NO_5$ & 0.00863   & 0.0222\\ \bottomrule 
	\end{tabular}}
\end{table}

\begin{table}  [H] 
	\centering
	\caption{GHG Emissions from different primary resources of BG \cite{HALDER2014444}}. 
	\label{Table:13}
	{\tabulinesep=0.5mm
		\begin{tabular}{lc}
			\toprule
			\textbf{Fuels}                & \textbf{ Emissions (Kg/Kg\hspace{1mm}fuel)}   \\ \midrule
			Rice husk            &  1.49 \\ 
			Bituminous coal      &  2.46 \\
			Natural gas          &  1.93 \\ \bottomrule  
	\end{tabular}}
\end{table}

\section{Conclusion}\label{sec6}

In this paper, a green energy-supplied off-grid heterogeneous cellular network along with IoT devices has been proposed and thoroughly analyzed. The feasibility of the proposed system has been evaluated through various key performance parameters such as per unit electricity generation cost, throughput, energy saving, energy efficiency, and carbon footprints. To find out the optimal criteria, wide-ranging simulations are accomplished using HOMER optimization software under different network conditions. From the simulation results, it is well established that locally available RESs are sufficient for powering the heterogeneous cellular network without any external support. The proposed system shows  its cost efficiency by endeavoring the minimum NPC and CoE. In addition, compared to other power supply solutions, the proposed system leads to negligible carbon emissions (0\% form solar PV and around 0.2\% form BG) that ensures improved eco-sustainability through green engineering solutions. Moreover, adequate energy storage devices and proper energy co-operation techniques enhance the system's reliability by maintaining the zero percent shortage/outage of energy. Quantitative results also reveal that the inter-BSs green energy-sharing mechanism system can save energy up to 27.98\%. Furthermore, the throughput and energy efficiency analysis guarantee the fulfillment of spectral efficiency and data rate requirement through efficient usage of energy. Finally, the proposed resource scheduling algorithm for the NRT application achieved a large amount of power-saving through permissible delay in comparison with the existing model. The novel eDRX and PSM assisted power-saving algorithm achieves a better battery lifetime for IoT devices. One future improvement of this work can be an autonomous droop controlled energy sharing policy. Additionally, for NRT and IoT applications, the proposed sleep mode algorithms can be optimized for specific applications. 

\bibliographystyle{IEEEtran}
\bibliography{ref}

\begin{IEEEbiography}[{\includegraphics[width=1in,height=1.25in,clip,keepaspectratio]{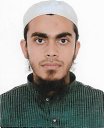}}]{KHONDOKER ZIAUL ISLAM} received the B.Sc. and M.Sc. degrees in electrical engineering from the Islamic University of Technology (IUT), Bangladesh. He is currently working as a Ph.D. student in the Murdoch University, Murdoch, Australia. He is an Assistant Professor with the Department of Electrical and Electronic Engineering, Bangladesh University of Business and Technology (BUBT),  Dhaka, Bangladesh. His current research interests include green communication, LPWAN in Agriculture, C-RAN, 5G cellular networks, radio resource management, and radio planning for cellular networks.
\end{IEEEbiography}

\begin{IEEEbiography}[{\includegraphics[width=1in,height=1.25in,clip,keepaspectratio]{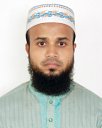}}]{MD. SANWAR HOSSAIN} received the B.Sc. degrees in electrical and electronic engineering from the Rajshahi University of Engineering and Technology, Rajshahi, Bangladesh, and the M.Sc. degree in electrical electronic and communication engineering from the Military Institute of Science and Technology, Dhaka, Bangladesh in 2010 and 2021 successively. From 2011 to 2015, he was a Lecturer, and currently, he has been serving as an Assistant Professor with the Department of EEE, Bangladesh University of Business and Technology, Dhaka, Bangladesh. His research interests include green energy, smart grid, and power system optimization. He is a graduate student member of IEEE and serves as a reviewer in many international journals
\end{IEEEbiography}

\begin{IEEEbiography}[{\includegraphics[width=1in,height=1.25in,clip,keepaspectratio]{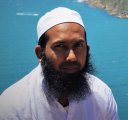}}]{B M Ruhul Amin} (S’17) received his B.Sc. Engg. and M.Sc. Engg. degrees in EEE from the Islamic University of Technology (IUT), Bangladesh. He is currently working as a Ph.D. student in the Macquarie University, Sydney, Australia. His current research topic is cyber secure control for future smart grid and his research interests include cyber security, control systems, smart grid, renewable energy integration, electric vehicles, and energy storage systems.
\end{IEEEbiography}

\begin{IEEEbiography}[{\includegraphics[width=1in,height=1.25in,clip,keepaspectratio]{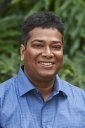}}]{Ferdous Sohel} (M’08-SM’13) received his PhD degree from Monash University, Australia, in 2009. He is currently an Associate Professor in Information Technology at Murdoch University, Australia. Prior to his joining Murdoch University, he was a Research Assistant Professor/Research Fellow at the University of Western Australia from 2008 to 2015. His research interests include computer vision, machine learning, pattern recognition, and digital agritech. He is a recipient of prestigious Discovery Early Career Research Award (DECRA) funded by the Australian Research Council. He is a winner of two WA State Govt. funded competitive grants on –shark hazard mitigation and digital pathology to improve cancer diagnosis. He is also a recipient of the VC’s Early Career Investigators award (UWA) and the best PhD thesis medal form Monash University. He has been serving as an Associate Editor of IEEE Transactions on Multimedia and IEEE Signal Processing Letters. He is a Member of Australian Computer Society and a Senior Member of the IEEE.
\end{IEEEbiography}

\EOD

\end{document}